\documentclass[preprint,showpacs,preprintnumbers,amsmath,amssymb,
nofootinbib,eqsecnum,12pt,floatfix]{revtex4-1}

\usepackage{graphicx}
\usepackage{comment}

\includecomment{versiona}

\newcommand{\mynote}[1]{}

\newlength{\dummysp}
\newcommand{\half}{\frac{1}{2}}

\newcommand{\beq}{\begin{eqnarray}}
\newcommand{\eeq}{\end{eqnarray}}
\newcommand{\nnn}{ \nonumber \\ }
\newcommand{\ddd}{ \nnn && }
\newcommand{\p}{{\partial}}
\newcommand{\e}{{\epsilon}}
\newcommand{\s}{{\sigma}}

\newcommand{\vev}[1]{{\langle #1 \rangle}}

\newcommand{\ord}[1]{{{\cal O}(#1)}}
\newcommand{\gappeq}{\mathrel{\rlap {\raise.5ex\hbox{$>$}}
{\lower.5ex\hbox{$\sim$}}}}
\newcommand{\lappeq}{\mathrel{\rlap{\raise.5ex\hbox{$<$}}
{\lower.5ex\hbox{$\sim$}}}}
\newcommand{\myref}[1]{(\ref{#1})}

\newcommand{\ben}{\begin{enumerate}}
\newcommand{\een}{\end{enumerate}}
\newcommand{\bit}{\begin{itemize}}
\newcommand{\eit}{\end{itemize}}

\newcommand{\Ncal}{{\cal N}}

\newcommand{\Ocal}{{\cal O}}

\newcommand{\Zbf}{{\bf Z}}

\newcommand{\muhat}{{\hat \mu}}

\def\[{\left [}
\def\]{\right ]}
\def\({\left (}
\def\){\right )}

\begin{document}

\title{Lattice sine-Gordon model}

\author{James Flamino}
\email{flamij@rpi.edu}
\author{Joel Giedt}
\email{giedtj@rpi.edu}
\affiliation{Department of Physics, Applied Physics and Astronomy,
Rensselaer Polytechnic Institute, 110 8th Street, Troy NY 12065 USA}

\date{March 16, 2018}

\begin{abstract}
We obtain nonperturbative results on the sine-Gordon model using the lattice field technique.
In particular, we employ the Fourier accelerated hybrid Monte Carlo algorithm for our
studies.  We find the critical temperature of the theory based on 
autocorrelation time,
as well as the finite size scaling
of the ``thickness'' observable used in an earlier lattice study by Hasenbusch et al.
We study the entropy, which is smooth across all temperatures,
supportive of an infinite order transition.  This system has
a well-known duality with the massive Thirring model, which can
play the role of a toy models for Montonen-Olive duality in $\Ncal=4$ super-Yang-Mills
theory, since it relates solitons to elementary field excitations.
Our research lays a groundwork for such study on the lattice.
\end{abstract}



\maketitle

\section{Introduction}
\begin{figure}
\begin{center}
\begin{minipage}{3in}
\includegraphics[width=3in]{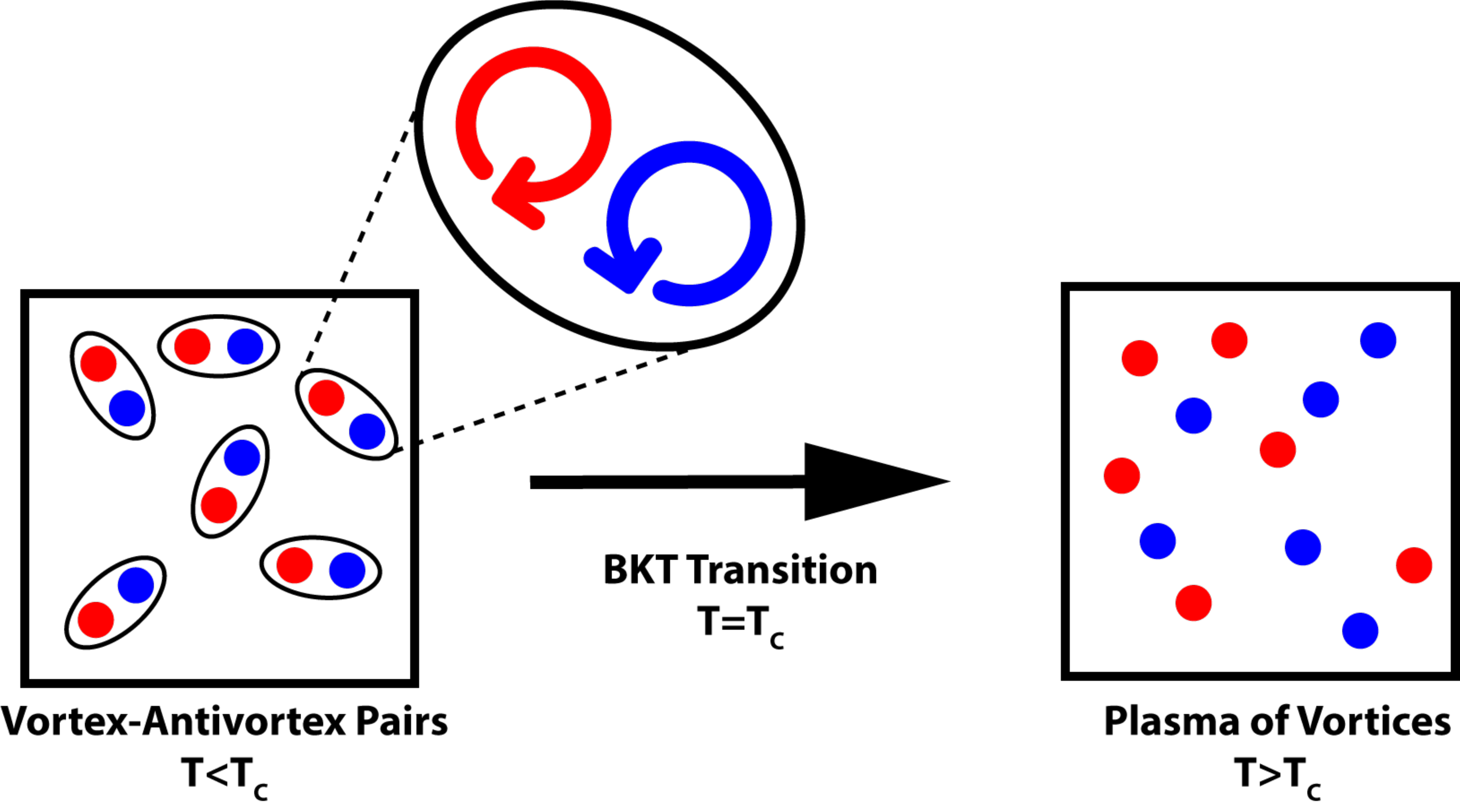}
\end{minipage}
\hfill
\begin{minipage}{3in}
\caption{Vortex binding and unbinding on the two side of the BKT
transition.  Figure taken from \cite{Giedt:2017teu}. \label{bindingvorts} }
\end{minipage}
\end{center}
\end{figure}

There were many important advances in theoretical physics in the early 1970s.
Among these was the discovery of phase transitions that were not characterized
by spontaneous symmetry breaking and long-range order.  The dynamics of vortices,
and the corresponding topological phase transitions, provided a way to have
critical phenomena while still satisfying the
theorems forbidding spontaneous symmetry breaking in two dimensions.  
The XY model and the two-dimensional (2d) Coulomb gas provide a system where
this physics of vortices and topological order can be studied.  They
also played an important role in the development of the
Wilsonian renormalization group.  The binding and unbinding of vortex -- anti-vortex pairs on
either side of the transition is shown in Fig.~\ref{bindingvorts}.

The Berezinskii-Kosterlitz-Thouless (BKT) transition \cite{Berezinskii:1971,Kosterlitz:1973xp} was 
originally formulated to describe the superfluid transition in two dimensions, such as the ${}^4$He thin
film.  It was subsequently applied to superconducting thin films, which are a sort of charged superfluid.
As a result, the supercurrents screen the fluctuations so care must be taken in attempting
to apply the BKT theory to this type of system.
The screening length is given by $\Lambda = \lambda^2/d$, where $\lambda$ is
the magnetic penetration depth and $d$ is the film thickness.  If the disorder is large,
then $\lambda$ is also large.  If in addition, the film is very thin, so that
$d$ is very small, $\Lambda$ can be large.  Then we can approximately neglect
the screening, and BKT can be applied.\footnote{See
for instance the review \cite{sgrvw1}.}  

The BKT transition can be contrasted with the Ginzburg-Landau
transition.  A key difference is the
absence of a conventional order parameter, though in the case of the
XY model one can measure vorticity to distinguish the two phases.  
But one particularly interesting feature of the XY model is that there is a
line of conformal fixed points, rather than a single temperature
at which the theory is critical.  Throughout this low temperature regime there
is algebraic ordering of the $O(2)$ spin fields; i.e., correlation functions yield
power laws of the separation between operators.  Thus one should view this
system as a family of conformal field theories (CFTs), since the
anomalous dimensions (critical indices) are continuously varying with
temperatures.  For this reason the XY universality class can be regarded
as a two-dimensional (2d) toy model for $\Ncal=4$ super-Yang-Mills (SYM),
which also has continuously varying anomalous dimensions for (composite)
operators depending on the gauge coupling $g$ (or more generally, the
complexified coupling which incorporates the $\theta$ angle).

In this article we will study the sine-Gordon model.  It
is the effective theory of the vortices in the XY model,
and has the Euclidean action:\footnote{Other
parameterizations are discussed in Appendix \ref{s:maps}.} 
\beq
S[\phi] = \frac{1}{t} \int d^2 x ~ \left\{ \half [ \p_\mu \phi(x) ]^2 - g \cos \phi(x) \right\}
\label{sgact}
\eeq
Here, $t$ is the ``stiffness.'' Interestingly, it corresponds to the inverse temperature of the corresponding
XY model.  The quantity $y=g/2t$ is the fugacity of vortices.
The small $g$, or small $y$, behavior of the sine-Gordon theory is well understood.
For $t > 8\pi$ (the low temperature regime of the XY model), the renormalization group (RG) flow (toward the infrared)
of $g$ is $g \to 0$.  We thus recover the theory
with long range correlations (algebraic ordering) in this part of the phase diagram.
By contrast, for $t < 8\pi$ (the high temperature regime of the XY model), the
flow of $g$ is $g \to \infty$, which leads to screening and an
absence of criticality.

Since $t > 8\pi$ corresponds to the low temperature
phase of the XY model, this is where the line of fixed points
will occur.  It is here that one has continuously varying critical
exponents, and conformal field theories will
describe the infrared physics.  Since this is a two-dimensional theory,
the conformal group is infinite dimensional, with generators $L_m$ satisfying the
well-known Virasoro algebra.  Note that this infinite dimensional symmetry algebra emerges at long distances,
associated with the flow $y \to 0$ of the renormalization group.
Thus it is challenging to obtain the composite field operators of the ultraviolet
theory that will correspond to the Virasoro generators.  This is because the
conformal symmetry is not a feature of the ultraviolet theory, due to the nonzero $y$.

As mentioned above, the temperature $t$ of the sine-Gordon
theory maps to an inverse temperature in the XY model.  
In the XY model, the correlation length $\xi$ approaches criticality
according to $\xi \sim e^{a/\sqrt{t_\text{red}}}$, where $t_\text{red}=(T-T_\text{BKT})/T_\text{BKT}$ is the
reduced temperature.  In the sine-Gordon theory this becomes $\xi \sim e^{b/\sqrt{8\pi-t}}$ for $t < 8\pi$,
where $\xi$ is the correlation length.  Thus we continue to have an essential singularity in the critical temperature/stiffness.
A result of this is that the transition is of infinite order, so that derivatives of the free
energy will be smooth functions.  We find results consistent with this in our
study of the entropy below, in Section \ref{s:ent}.

\begin{figure}
\begin{center}
\includegraphics[width=3in]{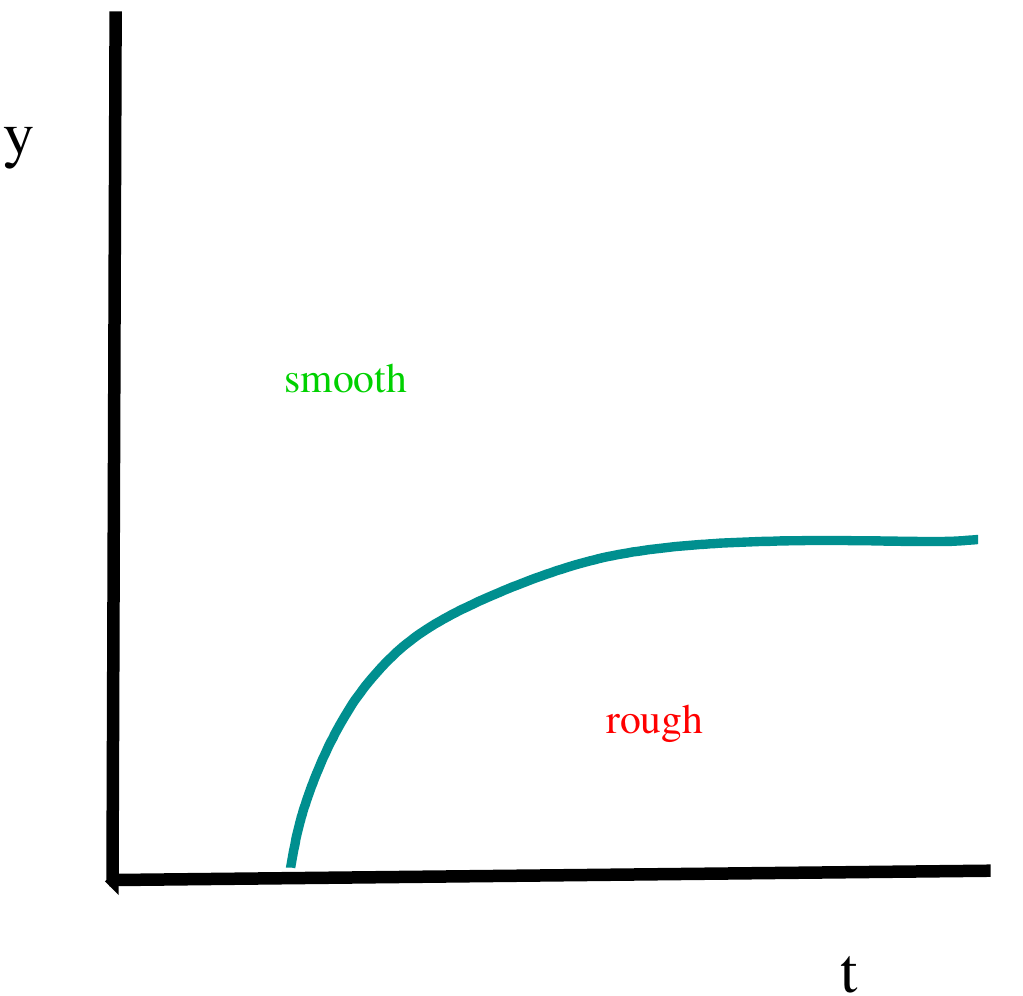}
\caption{Sketch of hypothetical phase boundary for sine-Gordon model.  \label{phasecurve} }
\end{center}
\end{figure}

The sine-Gordon model belongs to the family of solid-on-solid (SOS) models which have
critical behavior consistent with 
the BKT transition.  When one looks into the origin of the SOS models, one finds a roughening transition.
The correspondence between roughening transitions in crystal facets and the XY universality
class was first discussed in \cite{ChuiWeeks76,ChuiWeeks78}.
The variable $\phi(x)$ of the sine-Gordon model is interpreted as a height variable above a two-dimensional (2d)
surface---the facet of a crystal.  Below, we display figures that visualize this
picture from actual simulations; see Figs.~\ref{domain_fig1} and \ref{domain_fig2}.  
As a result, above the critical $t$, where $\phi(x)$ 
becomes highly nonuniform,
the sine-Gordon model describes the high temperature growth of a rough surface.  
There is a critical line $t_c(y)$, and the fugacity $y$ labels different types of crystals.
When $y \gg 1$, the cosine potential term in the action \myref{sgact} dominates and one is driven
to the uniform ground states where $\phi(x)$ is frozen to a multiple of $2\pi$.  Thus we
expect the curve $t_c(y)$ to tend to infinity as $y$ is increased, as shown in the
sketch, Fig.~\ref{phasecurve}.

It was mentioned above that the sine-Gordon theory is a toy model for ${\cal N}=4$ super-Yang-Mills
because it has a line of fixed points where nontrivial conformal field theories exist.  There is
an additional correspondence, namely the duality between solitons and elementary excitations.
In the present case, the sine-Gordon theory is known to be dual to the massive Thirring model,
with kink-like solitons in the sine-Gordon theory being dual to elementary fermions in the Thirring
theory \cite{Coleman:1974bu}.  A similar duality exists in ${\cal N}=4$ super-Yang-Mills:  on the Coulomb branch,
W-bosons are dual to 't Hooft-Polyakov monopoles, which are type of soliton.  Thus by
studying duality on the lattice in the context of the sine-Gordon model and Thirring model,
we can lay groundwork for the eventual study of electric-magnetic S-duality in ${\cal N}=4$
super-Yang-Mills.  This has a practical advantage in that simulating two-dimensional models
is far less costly in terms of computer resources.

In this article we explore the behavior of two useful observables in
a lattice formulation of the sine-Gordon model.  We use the method
of Fourier accelerated hybrid Monte Carlo, discussed in the next section.
We are able to identify the value of $t$ where the topological transition
occurs, and find that it agrees with previous determinations.  We also
provide some evidence that the transition is of infinite order.  Preliminary
aspects of a part of this study have previously appeared in \cite{Giedt:2017teu}.

We now summarize the content of this article.  In Section \ref{s:fa} we review
Fourier acceleration of hybrid Monte Carlo.   In Section \ref{s:2dcg} we review the
relation between the 2d Coulomb gas and the sine-Gordon model, which also
allows us to exhibit the fugacity expansion.  This is relevant to various
statements in later parts of the paper.  In Section \ref{s:sol} a few
words about the solitons of this model are given.  These are again well-known
facts.  After this we move on to our results, beginning in Section \ref{s:sim} where
we show our main simulation results and the thickness observable.  We find
results consistent with those obtained in an earlier study by Hasenbusch et al.~\cite{Hasenbusch:1994ef},
which used a cluster algorithm.  Next in Section \ref{s:ent} we show our explorations
of entropy, and find a picture that is consistent with an infinite order transition.
Section \ref{s:clu} takes a closer look at the domains that form when the barrier
height is significant.  We close the paper with Conclusions and a number of
Appendices that review some details related to earlier discussion, including
finer points of well-known results.  The appendices also discuss some of the
technical points of our implementation, and general musings about the fugacity
expansion.

\section{Fourier acceleration}
\label{s:fa}
In this section we briefly discuss the Fourier accelerated hybrid Monte Carlo (HMC) algorithm that is
the basis of our simulations \cite{Duane85,Ferreira93}.  The point this generalization of HMC is to reduce
autocorrelations between configurations of $\phi(x)$ that are produced in the
course of the simulation.  Provided this is successful, a shorter simulation can be used, while
producing equivalent results and uncertainties.  Previous work using Fourier
accelerated HMC includes \cite{Espriu:1997jh,Catterall:2001jg}.

At the beginning of the HMC simulation, the momentum field $\pi(x)$ is drawn
at random from a Gaussian distribution with unit variance.  This corresponds to
``integrating in'' a Gaussian field in the partition function, leading to the
``Hamilitonian''
\beq
H = \sum_x \bigg\{ \half \pi^2(x) - \frac{1}{2t} \phi(x) \Delta \phi(x) - \frac{g}{t} \cos \phi(x) \bigg\}
\eeq
where as usual the Laplacian is discretized by
\beq
\Delta \phi(x) = \sum_{\mu=1}^2 \p_\mu^* \p_\mu \phi(x) = \sum_{\mu=1}^2 \[ \phi(x+\muhat)
+ \phi(x-\muhat) - 2 \phi(x) \]
\eeq
Here, $\p_\mu$ is the forward difference operator in the $\mu$ direction,
$\p_\mu^*$ is the backward difference operator, and $\muhat$ is a unit vector in the
$\mu$ direction.  As usual, we work in lattice units, $a=1$.  The Hamiltonian $H$
is evaluated to obtain $H(0)$.  Next, the fields $\pi(x),\phi(x)$ are
evolved according to Hamilton's equations
\beq
\dot \phi(x) = \frac{\p H}{\p \pi(x)}, \quad \dot \pi(x) = - \frac{\p H}{\p \phi(x)}
\eeq
for a trajectory of length $\tau$, which we typically set to 1.  The numerical integration technique for this
evolution in fictitious time should be reversible and area-preserving, where area
refers to the functional integration measure $[d\pi(x) ~ d\phi(x)]$.  The standard
method is the leapfrog algorithm, which is what we use.  This integrator has step size $dt$,
and the Hamiltonian $H(\tau)$ at the end of the leapfrog trajectory will differ
from $H(0)$ due to $dt$ not being zero.  To correct for this, we apply the Metropolis accept/reject step
\beq
P_\text{acc} = \min(1,e^{-\Delta H}), \quad \Delta H = H(\tau) - H(0)
\eeq
to obtain a ``perfect'' algorithm, which will sample the functional integral with the
correct weight, the canonical distribution corresponding to $H[\pi(x),\phi(x)]$.

The Fourier acceleration is introduced into the leapfrog trajectory, where the
Fourier modes of $\phi(x)$ and $\pi(x)$ are integrated with a step size
\beq
dt(k) = dt/(\Delta(k) + m_\text{eff}^2)
\label{dtk}
\eeq
An equal number of steps $N_\tau = \tau/dt$ are taken for each mode $k$.
The Fourier transform of the force $-\p H/\p \phi(x)$ must also be used
in these equations.  In \myref{dtk}, $\Delta(k)$ is the Fourier transform
of $-\Delta(x,y)$:
\beq
\Delta(k) = \sum_{\mu=1}^2 4 \sin^2(\pi k_\mu / L_\mu), \quad k_\mu = 0,\ldots,L_\mu-1
\eeq
for an $L_1 \times L_2$ lattice.  By evolving the longer wavelength modes with a
larger step size, they are moved farther in configuration space.  This tends to reduce
autocorrelations, because it is precisely the long wavelength modes that
lead to long autocorrelation times.  

It is certainly not necessary to take the $k$ dependent step size $dt(k)$ to be
of the form \myref{dtk}.  A further study that we will conduct at a future
point is to generalize the choice of $dt(k)$ and then
optimize it using machine learning techniques.  The figure of merit that will
be maximized (i.e., the training goal) is the inverse of the integrated autocorrelation time.
Our current study that has the feature of being a first step in
extensive investigations of the Fourier
acceleration technique, using the sine-Gordon model as easily simulated toy model.

\section{Relation to the two-dimensional Coulomb gas}
\label{s:2dcg}
An important property of the SG model is its relation to the 2d Coulomb gas.
It is here that one can understand its dynamics in terms of a collection
of charges.  Concrete applications include the physics of defects in
liquid crystals, vortices in superconducting thin films, and vortices
in 2d superfluids, such as a monolayer of $^4$He on a substrate.  
Such features can also be realized in cold atom arrays.
It is worth noting that some of the most important properties of BKT physics
can be obtained from the multiple perspectives.  For instance the
renormalization group equations that were originally obtained by Kosterlitz \cite{KosterlitzRG74}
can also be obtained from the sine-Gordon model \cite{Amit80}.

The correlation functions that are the focus of this aspect of the sine-Gordon model
are not those of the elementary fields $\phi(x)$.  Rather, we work with what
high energy physicists would call ``composite operators,''
$e^{\pm i \phi(x)}$.  These have an interpretation as vortex and antivortex
creation operators.  

\subsection{Two-point function}
The basic two-point function to be considered is
\beq
G(x-y) = \langle e^{i \phi(x) } e^{-i \phi(y) } \rangle
\eeq
which corresponds to the amplitude for the creation of a vortex at $y$ and its
subsequent annihilation at $x$.
The action that determines this amplitude in the path integral has been given above in Eq.~\myref{sgact},
leading us to evaluate
\beq
&& G(x) = Z^{-1} \int [d\phi(z)] e^{-S_0(\phi)} e^{i [ \phi(x) - \phi(0) ] }
e^{(g/t) \int d^2 y \cos \phi(y)}
\ddd Z = \int [d\phi(z)] e^{-S_0(\phi)} e^{(g/t) \int d^2 y \cos \phi(y)}
\ddd S_0(\phi) = \frac{1}{2t} \int d^2 x ~ [ \p_\mu \phi(x) ]^2
\eeq
We proceed by expanding the exponential
$$
e^{(g/t) \int d^2 y \cos \phi(y)}
$$
in powers of $g/t$:
\beq
&& \sum_{n=0}^\infty \frac{1}{n!} \( (g/t) \int d^2 y ~ \cos \phi(y) \)^n
= 1 + \frac{g}{2t} \int d^2 y ~ \( e^{i \phi(y)} + e^{-i \phi(y) } \)
\ddd + \frac{g^2}{8t^2} \int d^2 y_1 ~ d^2 y_2 ~ \big( e^{i [ \phi(y_1) + \phi(y_2) ] }
+ e^{-i [ \phi(y_1) + \phi(y_2) ] }
\ddd + e^{i [ \phi(y_1) - \phi(y_2) ] }
+ e^{-i [ \phi(y_1) - \phi(y_2) ] } \big) + \ldots
\eeq
In the correlation function, only the terms with an equal number of plus signs
and minus signs in the exponent will survive.\footnote{It turns out that $n$ must be even because of a charge neutrality condition
requiring cancelations of signs, due to the logarithmic divergence of the
long distance 2d Coulomb potential (see e.g.~\cite{ZinnJustin}).}  This further implies that only
even powers of $g$ will contribute ($n$ even).
The interpretation of the $n$th term in this sum 
\beq
\prod_{i=1}^n e^{i \e_i \phi(y_i)}, \quad \e_i = \pm 1, \quad \sum_{i=1}^n \e_i = 0, \quad n \in 2 \Zbf
\eeq
is the virtual pair production of $n/2$
vortex-antivortex pairs that interact with the vortex at $y$ and antivortex at $x$
as well as with each other.
  The connection between the 2d Coulomb
gas, the interpretation in terms of vortices of the XY model, and the sine-Gordon model
has been explained in many places; see for instance \cite{samuel,sgrvw1,ZinnJustin,Mudry}.
One thing that is interesting is that the temperature $t$ of the sine-Gordon (roughening) model
maps to an inverse temperature in the corresponding 2d Coulomb gas model.  Hence there is
a sort of $t \to 1/t$ duality between these equivalent descriptions,
reminiscent of the Kramers-Wannier duality in the 2d Ising system  or
the strong-weak coupling duality $g \to 1/g$ (S-duality) in $\Ncal=4$ super-Yang-Mills.
Renormalizing the composite operators\footnote{See for instance Section 32.1 of Zinn-Justin \cite{ZinnJustin}
for further details.  Whereas the free theory is obviously finite and free of UV divergences, the
same is not true of composite operators in that theory.  They can have (trivial) short-distance
singularities, as is the case of the operator discussed here.}
\beq
e^{i \e_i \theta(x)} = \zeta \[ e^{i \e_i \theta(x)} \]_{\text{ren.}},
\quad
\zeta = ( \Lambda/\mu )^{-t/4\pi }
\eeq
we obtain renormalized averages of these operators in the theory with action $S_0$ (i.e., the free theory)\footnote{Details of this calculation are given in Appendix \ref{ftprop}.}
\beq
\left\langle \prod_{i=1}^n e^{i \e_i \phi(x_i)} \right\rangle_{\text{ren.}}
= \prod_{i < j} ( \mu |x_i - x_j| )^{\e_i \e_j t/2\pi}
\label{zeroth}
\eeq
From this we find that
\beq
&& \left\langle e^{i [ \phi(y) - \phi(z) ] } \right\rangle_{\text{SG}}
\ddd = Z_{\text{SG}}^{-1} \sum_{n \in 2\Zbf} \frac{g^n}{2^n t^n n!} \int d^2x_1 \cdots d^2x_n
\sum'_{\{\e\}} \prod_{i<j} ( \mu |x_i - x_j| )^{\e_i \e_j t/2\pi}
\label{examexp}
\eeq
Here $i,j = 0,\ldots,n+1$ with $x_0=y$, $x_{n+1}=z$ and $\e_0=1$, $\e_{n+1}= -1$ so that they are
not summed over, hence the primed summation symbol.  $Z_{\text{SG}}$ is the same expression but
without the $i,j = 0$ and $n+1$ terms (vacuum diagrams only). I.e.,
\beq
Z_{\text{SG}} = \sum_n \frac{g^n}{2^n t^n n!} \int d^2x_1 \cdots d^2x_n
\sum_{\{\e\}} \prod_{i<j} ( \mu |x_i - x_j| )^{\e_i \e_j t/2\pi}
\eeq
with $i,j = 1,\ldots,n$ only.

Note that there are singularities when $\e_i \e_j < 0$ and $x_i \to x_j$
under the integration.  It is not an integrable singularity for $t \geq 4\pi$.
So it is not the case that we have rendered the correlation function finite
to all orders simply by renormalizing the composite fields by the factor $\zeta$.
All that has been accomplished is to make \myref{zeroth} finite, but the
subsequent integration of this expression in the sine-Gordon model
leads to further difficulties.\footnote{Although we have emphasized
the short distance singularities $x_i \to x_j$ at the beginning
of this paragraph, there will also be long distance divergences
coming from $\e_i \e_j > 0$ if we work in infinite 2d space. }
This is one reason to perform the lattice
discretization, because all quantities are under control and finite.  

Hard-core repulsion can be inserted into the above formula,
thereby avoiding $x_i = x_j$.  Of course, this does not follow
from the path integral of the sine-Gordon model directly,
but is an add-on to control the singularity and model the
regulation of this infinity in the physical system.
The long distance singularity that occurs from infinite $|x_i-x_j|$
can also be moderated by periodic boundary conditions
or some other finite system size regulation, which is again
a matter of inserting physical constraints on the
mathematical calculation in order to avoid the bad behavior.

\subsection{Observations about renormalization}
It may be that we would like to renormalize the theory, absorbing the
UV divergences in the low energy parameters.
To accomplish this in renormalized perturbation
theory, we would cancel these infinities with counterterms; or, equivalently,
we would rescale the bare fields and re-express the bare parameters
in terms of long distance quantities.  The problem
is whether redefinitions of $t$ and $g$, together with $\phi \to \sqrt{Z} \phi$,
suffices.  After all, classically, the
mass dimension of $\phi$ is $d_\phi = \half (d-2) = 0$,
so we can write down an infinite number of counterterms in the renormalizable
Lagrangian.  The potential $(g/t) \cos (\phi(x))$ is not unique according
to this power-counting based around the Gaussian fixed point.  However,
the field $\phi(x)$ will acquire an anomalous mass dimension, as will
operators built upon it; so we expect that only a finite number of operators
will be relevant or marginal once quantum effects are taken into account.
Certainly the observed universality of XY type models suggests this.\footnote{For
instance, including next-to-nearest neighbor couplings will shift the critical
value of the XY stiffness $K$, but not the critical exponents.  }

One constraint on the renormalized theory is the invariance of the
bare theory under the shift symmetry $\phi(x) \to \phi(x) + 2\pi$.
Certainly this forbids a host of terms in the potential.  However,
it allows for terms of the form $\cos[p \phi(x) + \alpha]$ where
$p$ is any integer and $\alpha$ is a real number.

The way that we would
normally proceed is to enumerate all of the one-particle irreducible (1PI) functions and vertex functions
that have a non-negative superficial degrees of divergence.  That would tell
us how many counterterms we need.  But since we are working in position
space (due to the composite operators and cosine potential) and have integrals over $x$, it is not clear how to do the counting
in a perturbation theory structured around the fugacity expansion.  
Also, this is in terms of composite fields, and we are not working with vertices
of elementary fields.  
In fact, the casual observer will not immediately see a diagramatic expansion
in the above formula.  So the usual tools seem to be off the table.  We will address this
conundrum in analysis that follows.

From the work of \cite{Hal79} we know that in superconducting thin films,
the behavior can be modeled as a superposition of BKT fluctuations and
Ginzburg-Landau (GL) fluctuations.  In our sine-Gordon model we will be
spared of this complication, because it only describes the vortex
dynamics of the XY model.  However it is interesting to think about how
this feature might be incorportated into the effective field theory, and
we will have some further comments on this direction of research in
the conclusions.

\section{Solitons}
\label{s:sol}
The minima of the potential $V(\theta)$ occur at
\beq
\phi(x) = 2 \pi n, \quad n \in \Zbf
\eeq
The solitons occur for static configurations
\beq
\frac{\p \phi(t,x)}{\p t} = 0
\eeq
Hence $\phi(t,x) = \vartheta(x)$.  The topologically nontrivial solitons asymptote to different vacua:
\beq
\lim_{x \to -\infty} \vartheta(x) \to 2 \pi n_-, \quad \lim_{x \to \infty} \vartheta(x)  \to 2 \pi n_+, \quad n_- \not= n_+
\label{asysol}
\eeq
We are particularly interested in the configurations that are minima of the configuration energy
\beq
E = \frac{1}{t} \int_{-\infty}^\infty dx ~ \left\{ \half (\p_x \vartheta)^2 + V(\vartheta) \right\}
\eeq
subject to the boundary conditions \myref{asysol}.
These are solutions to the static equations of motion (saddle point equation in this Euclidean formulation)
\beq
\frac{\delta E}{\delta \vartheta(x)} = 0  \quad \Rightarrow \quad
\p_x^2 \vartheta = \frac{\p V}{\p \vartheta}
\eeq
The solutions are well known; they are given for instance in the wonderful review of topological
solitons by 't Hooft, Chapter 4.1 of \cite{tHooftspell}.  For instance, in the case
of $n_- = 0$ and $n_+ = 1$, the soliton is described by the first branch of
\beq
\vartheta(x)= 4 \arctan e^{mx}, \quad m^2 = \frac{g}{t}
\eeq
Of course in our dynamical configurations this will not be exactly true (until we cool),
and so we would like to have a way of measuring the mass $m_{\text{eff}}$ of the
soliton from such non-extremal configurations.  We will explore this in future work.

On the lattice we have an $S^1 \times S^1$ geometry, so the asymptotic
behavior described above does not make sense.  However, we can still have
nontrivial topology by incorporating twisted boundary conditions in
the spatial direction $x=x_1$:
\beq
\theta(t,x+N_x a) = \theta(t,x) + 2 \pi n, \quad n \in \Zbf
\label{twbcs}
\eeq
This is also left to future work.

\section{Simulation results}
\label{s:sim}
As we approach the transition temperature, an increasing
number of degrees of freedom participate in the low energy theory,
as can be seen by the onset of ``algebraic ordering,'' corresponding
to a divergent correlation length.  This makes the fugacity
expansion unreliable, and in fact is the reason that
renormalization group methods were employed to understand
the transition.  We therefore turn to a numerical study of the model
where all features are regulated by the lattice's UV and IR cutoffs.

\subsection{$\phi$ distribution and vacuum degeneracy}
\begin{figure}
\begin{center}
\includegraphics[width=5in]{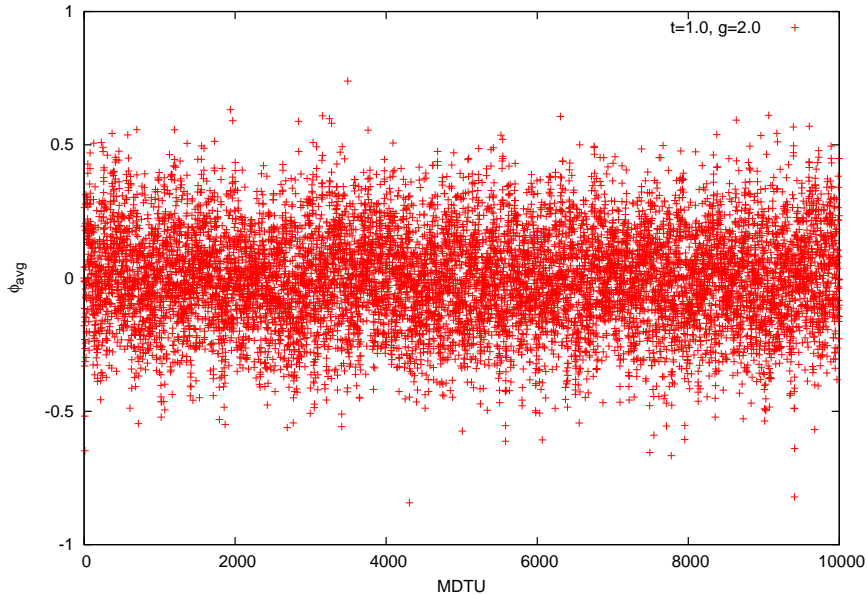}
\caption{The average value of $\phi$ on a $4 \times 4$ lattice as a function
of simulation time, for $t = 1.0$ and $g=2.0$.  It can be seen that the field
fluctuates about the vacuum it was started in, $\phi(x)=0 \; \forall \; x$.
The barrier to tunneling to other vacua is too great for the HMC to allow it.
\label{avg_phi_10}}
\end{center}
\end{figure}

\begin{figure}
\begin{center}
\includegraphics[width=5in]{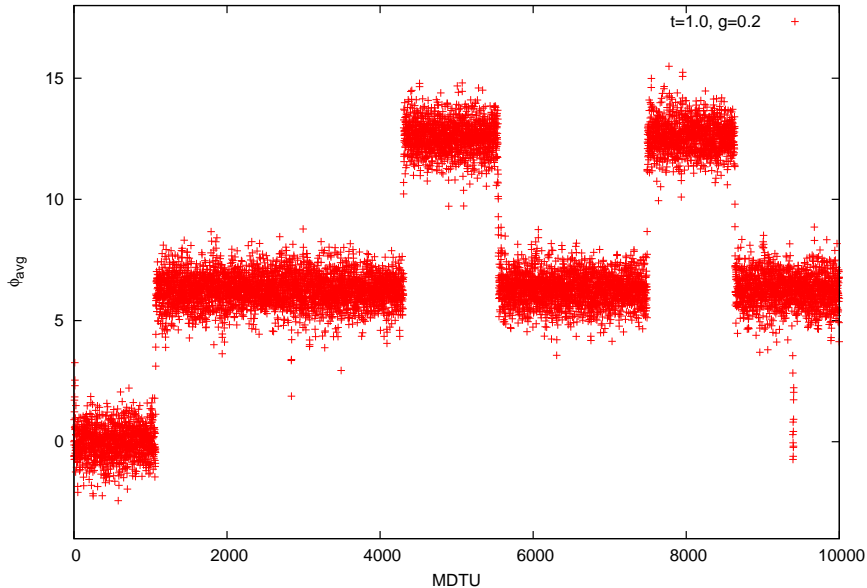}
\caption{The average value of $\phi$ on a $4 \times 4$ lattice as a function
of simulation time, for $t = 1.0$ and $g=0.2$.  I.e., a barrier
height 10 times lower than in Fig.~\ref{avg_phi_10}.  It can be seen that the field
now tunnels between vacua $\phi(x)= 2 \pi n \; \forall \; x$, $n \in \mathbb{Z}$.
With $g=0.2$, the barrier to tunneling to other vacua is low enough for the Fourier accelerated HMC to allow it.
Because the lattice is small, the entire field configuration moves coherently between
the vacua.
\label{avg_phi_02}}
\end{center}
\end{figure}

\begin{figure}
\begin{center}
\includegraphics[width=5in]{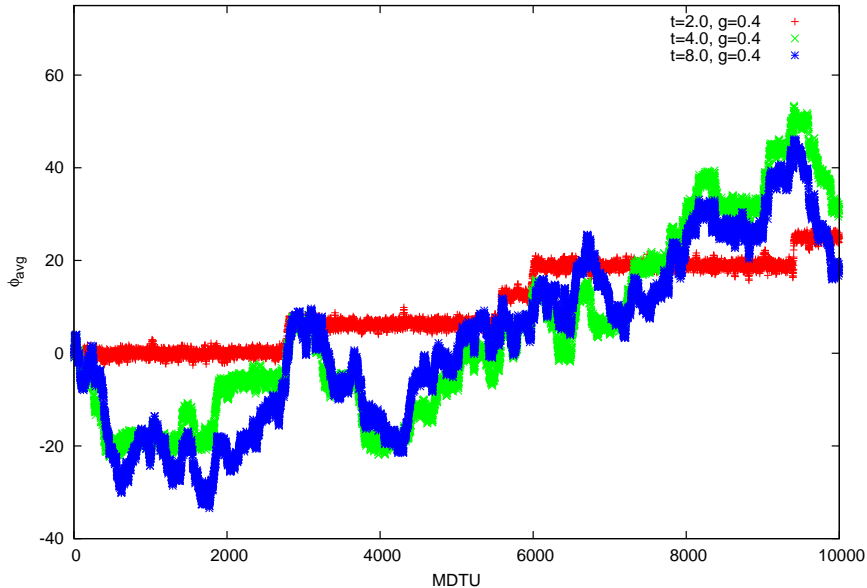}
\caption{The average value of $\phi$ on a $4 \times 4$ lattice as a function
of simulation time, for three different values of $t$ and $g=0.4$.  It can be seen that 
at smaller $t$ the field rarely tunnels between vacua $\phi(x)= 2 \pi n \; \forall \; x$, $n \in \mathbb{Z}$, 
whereas for larger $t$ the tunneling is fairly rapid.  This is because the height
of the barrier is proportional to $1/t$ and the strength of fluctuations is proportional to $\sqrt{t}$.  
\label{avg_phi_byt}}
\end{center}
\end{figure}

\begin{figure}
\begin{center}
\includegraphics[width=5in]{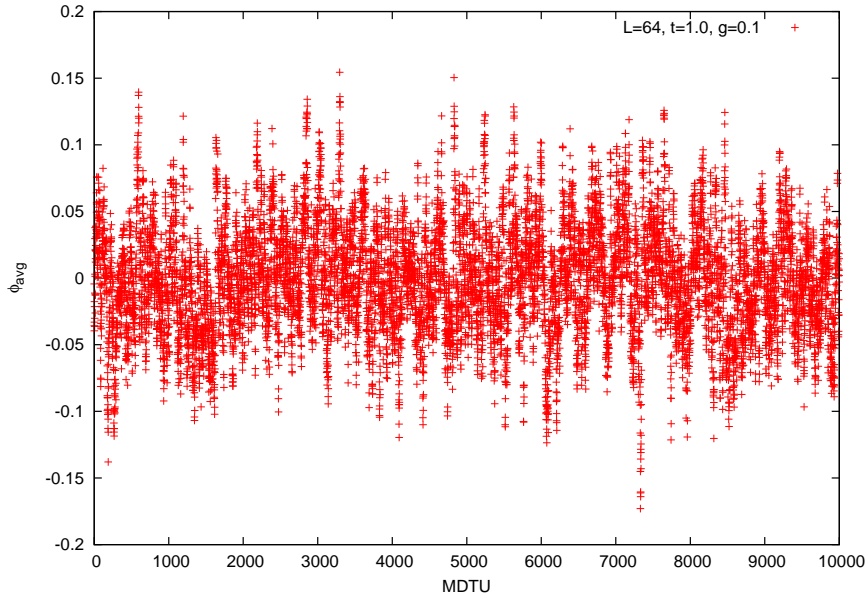}
\caption{The average value of $\phi$ on a $64 \times 64$ lattice as a function
of simulation time.  Although the couplings are the same as in Fig.~\ref{avg_phi_02},
the lattice has 256 times more sites, opening up the possibility of forming domains of the different
vacua.  If that was the case, they would tend to cancel in the average of $\phi$,
leading to values close to zero.  Thus a finer-grained study of the lattice
configurations is necessary in order to resolve what is going on.  In fact, 
Fig.~\ref{nodom} shows that this is not what is occuring.  Rather, the entire
lattice collapses almost immediately into a single domain, and remains there,
in contrast to what happens on the $4 \times 4$ volume of Fig.~\ref{avg_phi_02}.
\label{avg_phi_64_01}}
\end{center}
\end{figure}

\begin{figure}
\begin{center}
\includegraphics[width=5in]{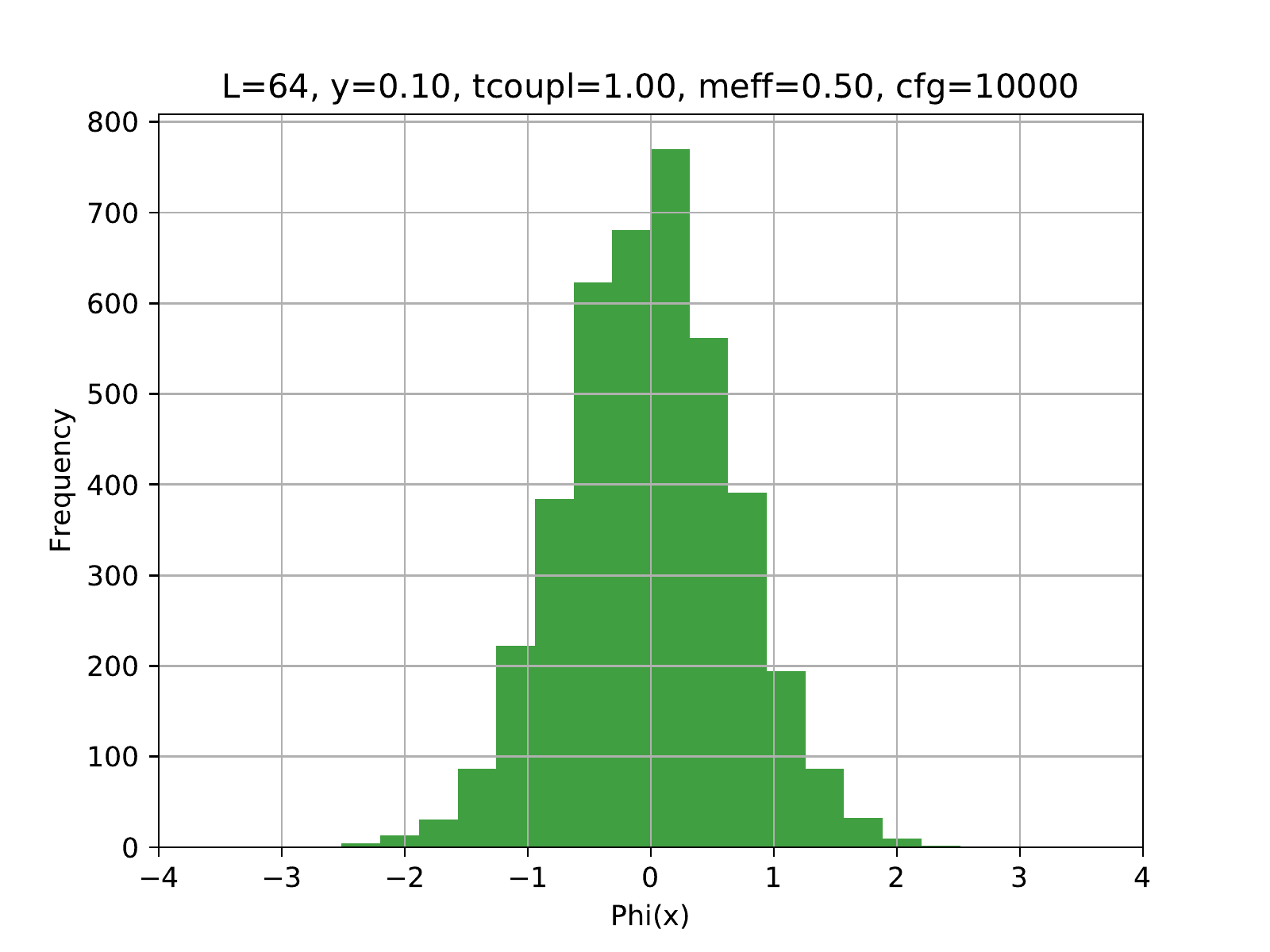}
\caption{Histogram of the $\phi$ value
for a simulation with lattice size $L=64$, 
temperature $t=1.0$, coupling $g=0.2$ ($y=0.1$), 
for configuration number 10,000 in the simulation, 
showing that it is centered
around one vacuum config, and not really forming domains. 
This answers the question that was raised in the caption
of Fig.~\ref{avg_phi_64_01}.
\label{nodom}}
\end{center}
\end{figure}

As we have seen in the previous
section, the form of the potential $V(\phi) \sim \cos(\phi)$ indicates that there will be
an infinity of degenerate vacua, $\phi \sim 2\pi n$, $n \in \mathbb{Z}$.  We
expect to see this feature in the simulations, although the tunneling 
reflects the algorithm in addition to the physical barrier between these
vacua.\footnote{In fact, it is an interesting problem to design
an HMC-type algorithm that easily moves over these barriers,
because of the analogy to topological charge
transitions in QCD.  In that context, simulations
close to the continuum limit with chiral lattice fermions are facing serious problems with
non-ergodicity, in that they fail to sample all topological sectors.  Perhaps it
would be easiest to study this problem in the SG model first.}  
In Fig.~\ref{avg_phi_10} we show the history of the average 
\beq
\phi_\text{avg} = \frac{1}{V} \sum_x \phi(x)
\eeq
where $V = L_0 L_1$ is the simulation volume.  Time along the bottom
axis is the simulation time, measured in molecular dynamics time units (MDTU).
It can be seen that for $t=1.0$ and $g=2.0$ (fugacity of $y=1.0$), no tunneling between the
various vacua occurs; here, the simulation was initiated at $\phi=0$.
Clearly the barrier is too large to allow HMC to move into other vacua.
In Fig.~\ref{avg_phi_02} we reduce the barrier ten-fold, setting $g=0.2$
(fugacity of $y=0.1$),
and now there is significant migration between the vacua.  Of course since
the barrier height is inversely proportional to $t$, the effective
tunneling rate will also depend on this quantity.  In Fig.~\ref{avg_phi_byt}
we show what occurs for three different values of $t$, all with $g=0.4$.
As per expectations, tunneling is significantly suppressed by decreasing $t$.
Fig.~\ref{avg_phi_64_01} shows a much larger lattice, $64 \times 64$ ($L=64$),
where it is possible that domains are forming, consisting of the different
vacua.  In this case the different vacuum values of $\phi$ would tend to
cancel, giving a result that fluctuates about zero.  
It can be seen in Fig.~\ref{nodom} that domains have not formed,
but rather on the larger lattice it is simply stuck near the $\phi=0$
vacuum, rather than tunneling.  Apparently, on the larger volumes, tunneling
is more difficult due to a tendency toward coherence across the
entire lattice.

\subsection{Thickness}
In \cite{Hasenbusch:1994ef}, Hasenbusch et al.~introduce the {\it thickness:}
\beq
\s^2=\frac{1}{V^2}\sum_{x,y} \vev{ (\phi(x)-\phi(y))^2 }
\eeq
In this formula, $V = L \times L$ is the system size, ``volume.''
They show that this is a useful observable for identifying the critical regime.
We will use it here, and find results consistent with theirs.
The thickness is a measure of the roughness, on average, for a given parameter pair $(t,y)$.
For example, if the entire lattice sits in
a single domain, with small fluctuations, then $\s$ will be small.
However, domain walls will contribute a nonzero result
even in the absence of fluctuations, so ground state disorder will
increase the thickness observable.

\begin{figure}
\begin{center}
\includegraphics[width=5in]{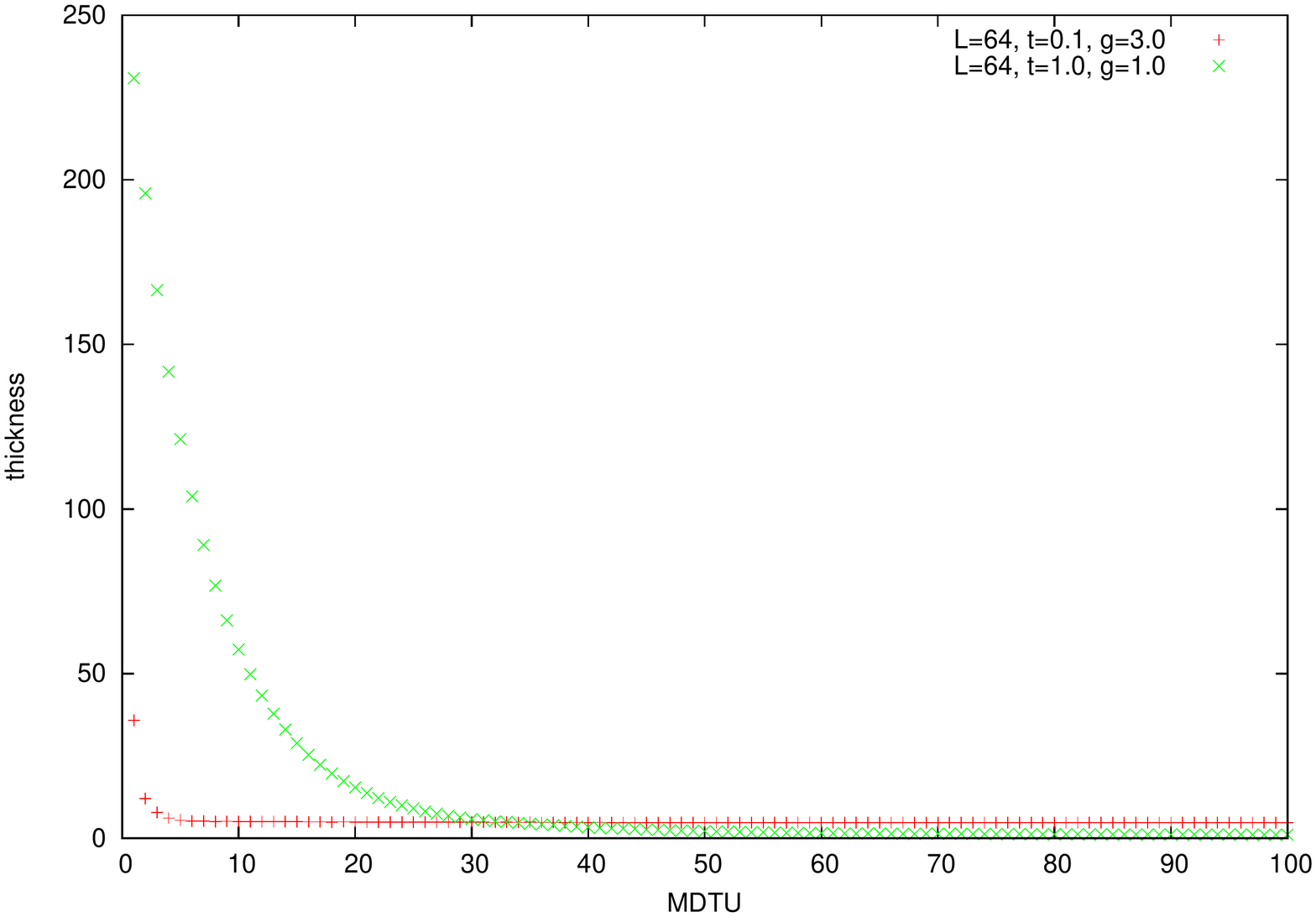}
\caption{The thickness on a $64 \times 64$ for two
different choices of $t$ and $g$, as function of
simulation time, beginning with the random start
described in the text.
For the case of a steep potential, $y=g/2t=15$, the fields
rapidly settle into minima and the thickness becomes small,
as it only gets contributions from domain walls between the minima.
Note however, that ultimately the thickness is largers because that
corresponds to small $t=0.1$, whereas the $t=1.0$ data has much
larger fluctuations.  This is because the first term in the action,
the kinetic term, is the one that determines the size of fluctuations
about any given classical vacuum.
\label{thick64}}
\end{center}
\end{figure}

In order to not bias toward a particular ground state, unless otherwise stated we begin all of our
simulations in studies of thickness with a
random start, as in Fig.~\ref{thick64}, which has $\phi \in [-20,20]$ uniformly 
distributed.

\begin{figure}
\begin{center}
\includegraphics[width=5in]{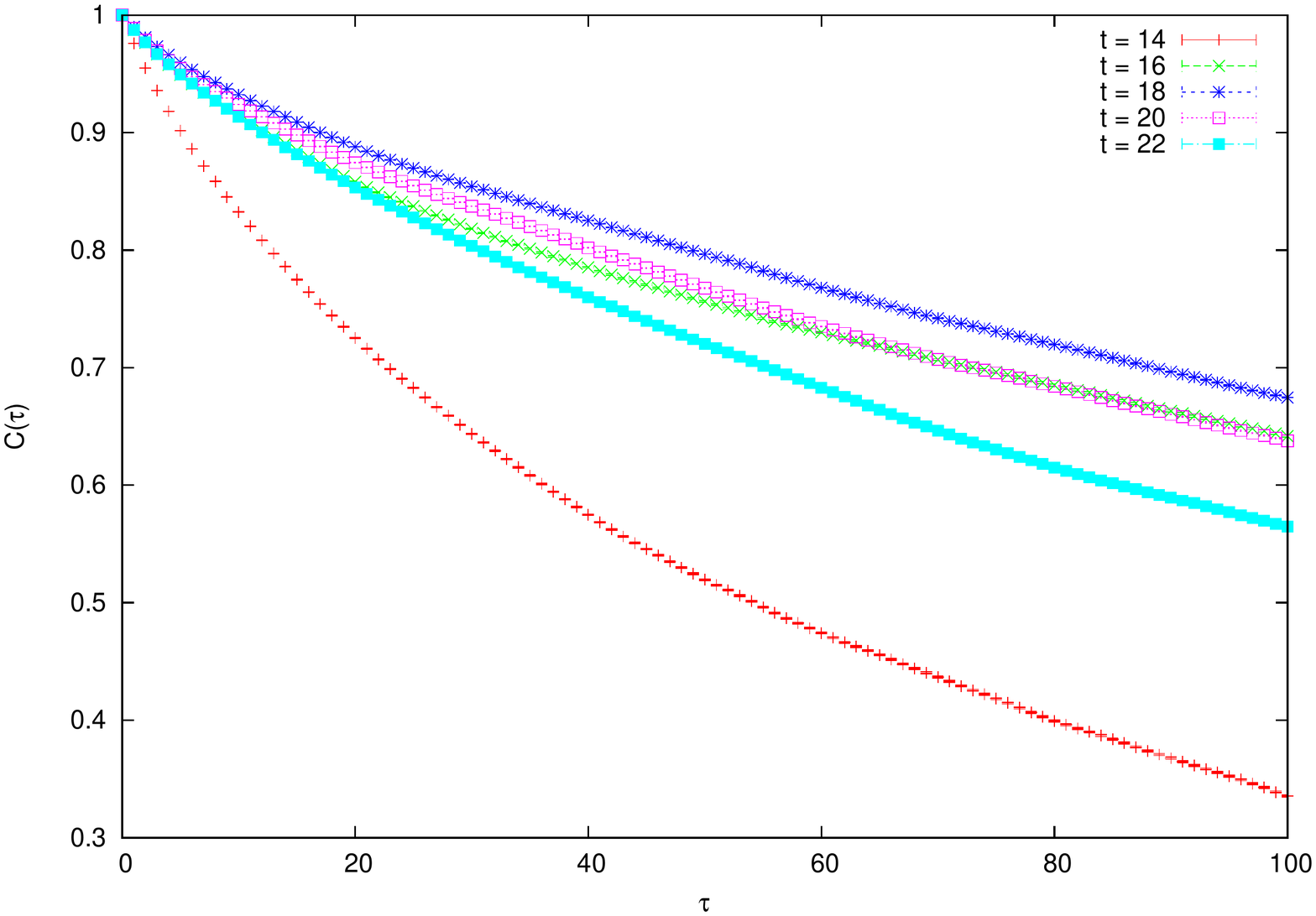}
\caption{The autocorrelation in the thickness observable
on a $64 \times 64$ lattice for $y=0.1$, and $t$ that
bracket the phase transition.  This figure shows short time scales,
with accompanying figures showing longer
time scales.  It can be seen that for $t \approx t_c \approx 18$,
the autocorrelation is the greatest, as is to be
expected.  Figure previously appeared in \cite{Giedt:2017teu}.
\label{acthick_short}}
\end{center}
\end{figure}

\begin{figure}
\begin{center}
\includegraphics[width=5in]{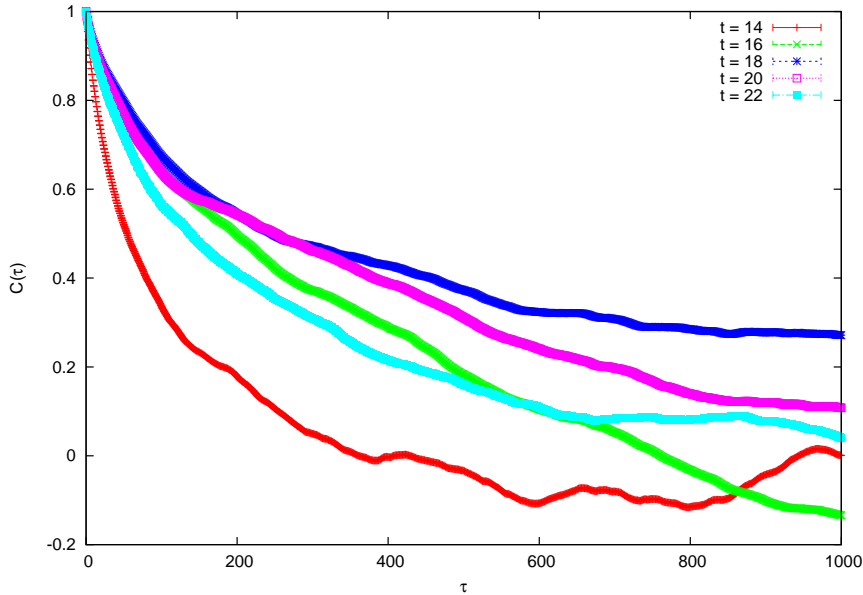}
\caption{The autocorrelation in the thickness observable,
showing longer time scales than Fig.~\ref{acthick_short}.
It can be seen that even after 1,000 updates, autocorrelations
are still significant for temperatures close to the
critical temperature of $t_c \approx 18$.
Figure previously appeared in \cite{Giedt:2017teu}.
\label{acthick_long}}
\end{center}
\end{figure}

\begin{figure}
\begin{center}
\includegraphics[width=5in]{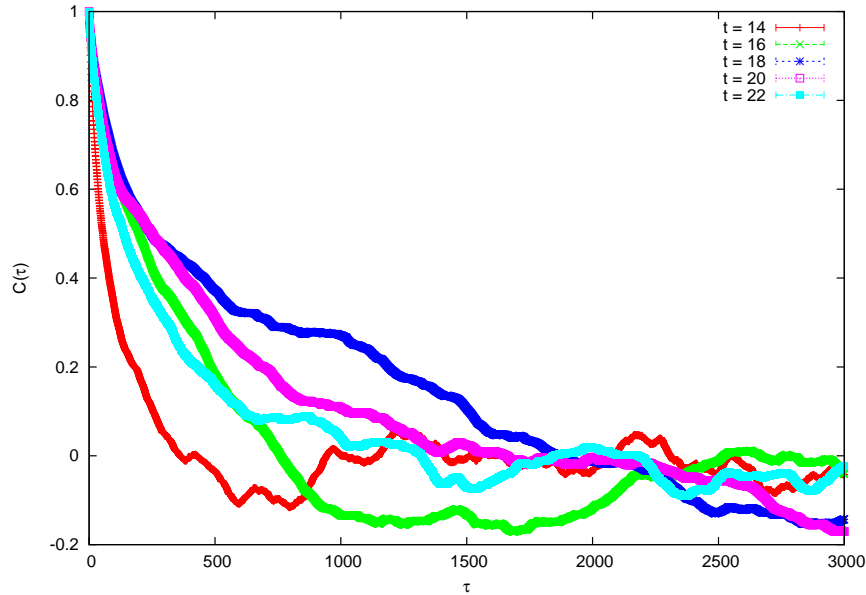}
\caption{The autocorrelation in the thickness observable,
showing very long time scales, compared to Figs.~\ref{acthick_short}
and \ref{acthick_long}.  It can be seen that by $\ord{3000}$ time
steps all memory of the initial configuration has vanished.
Figure previously appeared in \cite{Giedt:2017teu}.
\label{acthick_vlong}}
\end{center}
\end{figure}

Before taking averages and estimating uncertainties, one should study
and characterize the autocorrelation of the thickness observable.
This is because we need several autocorrelation times in
order to thermalize, and we need to know the
separation between statistically independent samples---where the
Monte Carlo simulation is effectively a Markov process.
The autocorrelation for any observable $\Ocal(t)$, where
$t$ here is the simulation time (measured in MDTUs), is
given by
\beq
C(t) &=& \frac{1}{{\cal N}} \frac{1}{N-t} \sum_{t' = 0}^{N-t-1} ( \Ocal(t'+t)-\vev{\Ocal} ) ( \Ocal(t')-\vev{\Ocal} ) \nnn
{\cal N} &=& \frac{1}{N} \sum_{t=0}^{N-1} (\Ocal(t) - \vev{\Ocal})^2,
\quad \vev{\Ocal} = \frac{1}{N} \sum_{t=0}^{N-1} \Ocal(t)
\eeq
and $N$ is the total number of time steps in the simulation,
$t = 0, 1, \ldots, N-1$.  In the
present case,
\beq
\Ocal = \frac{1}{V^2} \sum_{x,y} ( \phi(x) - \phi(y) )^2
\eeq
In Figs.~\ref{acthick_short}, \ref{acthick_long} and \ref{acthick_vlong}
we show short, long and very long time scales.  It can be
seen that there is an initial rapid decay, but that
a longer time component also contributes.  In
fact it takes $\ord{10^3}$ or more updates to obtain a completely
independent configuration.  These results are for $y=0.1$
with $t$ values that bracket what will turn out to be the
critical temperature, $t_c \approx 18$.  In fact, it can
be seen that $t \approx t_c$ yields the longest
autocorrelation times, which is to be expected.  This is
because critical fluctuations, which have very long
wavelength, lead to significant slow-down in typical
Monte Carlo algorithms.  We see that the Fourier acceleration
has not been entirely effective in alleviating this
critical slowing down.  Amusingly, monitoring the 
autocorrelation can be a surprisingly
good way to locate the critical temperature.

\begin{figure}
\begin{center}
\includegraphics[width=4in]{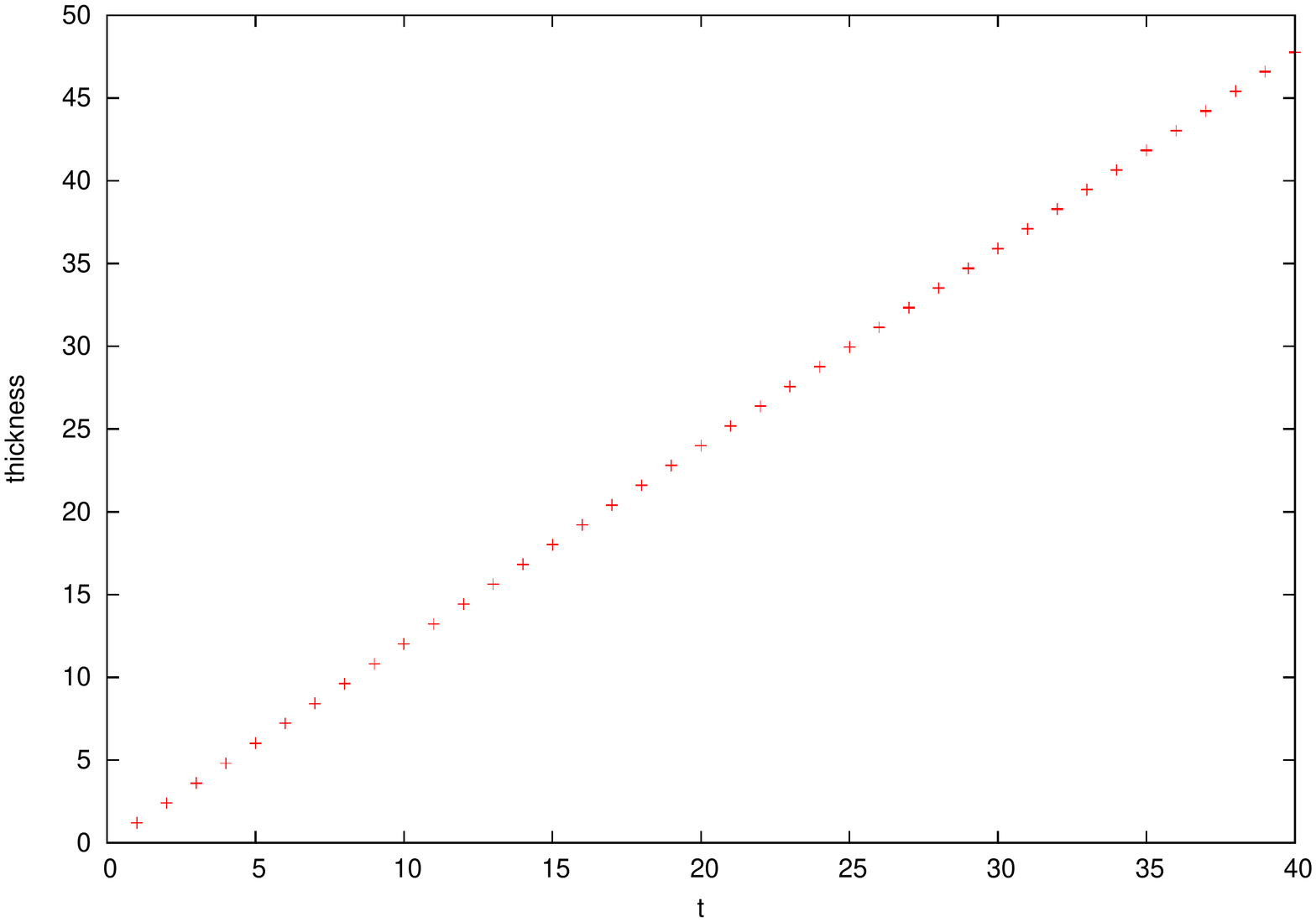}
\caption{Thickness versus $t$ for $y =0.1$ for an $L=32$ lattice. It can be seen
that it is a linear function of $t$ for this relatively weak value of $y$. \label{thvty010}}
As will be seen in subsequent figures, a more interesting behavior occurs for $y > 1$.
\end{center}
\end{figure}

\begin{figure}
\begin{center}
\includegraphics[width=4in]{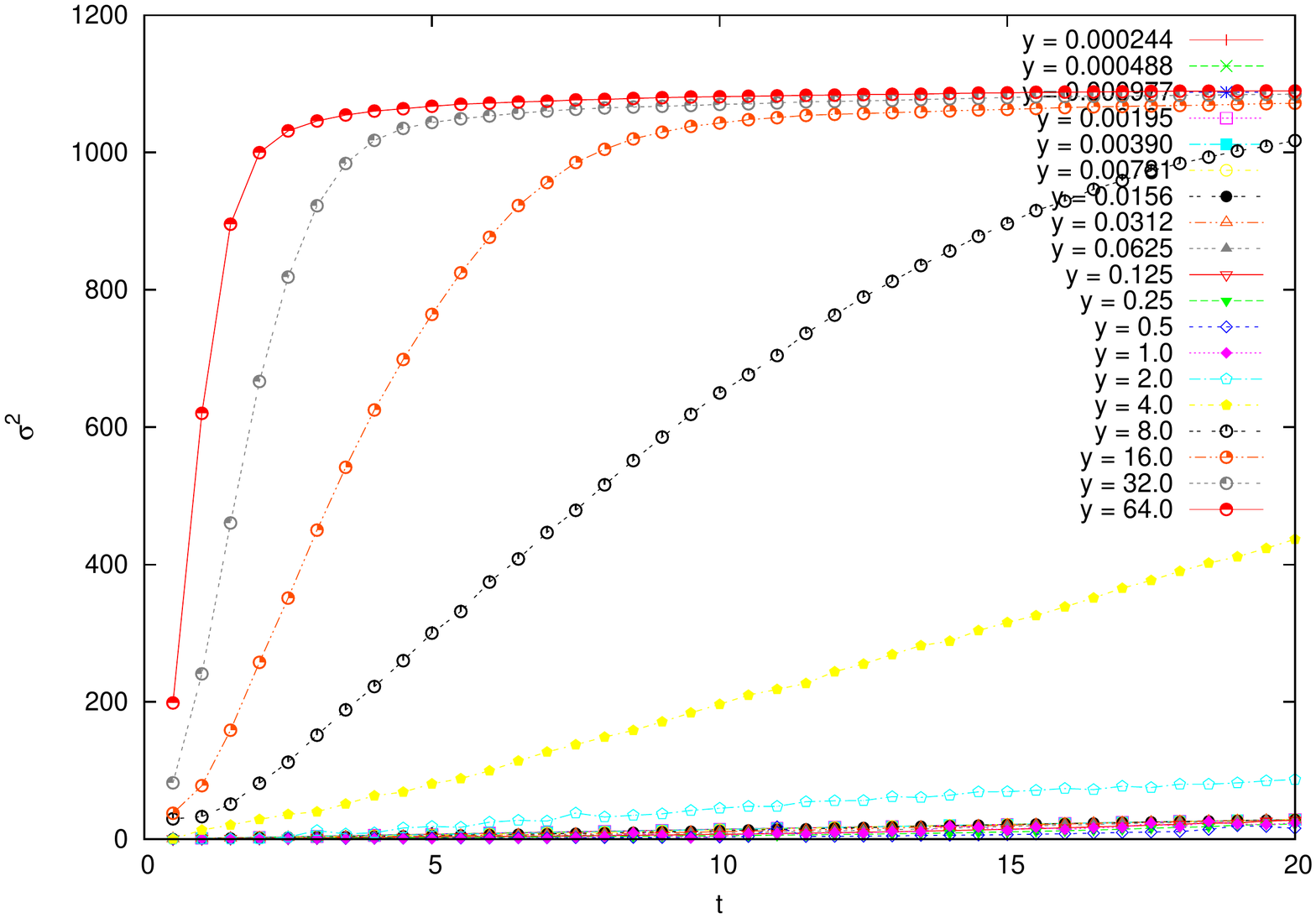}
\caption{Thickness versus $t$ for various $y$ for an $L=64$ lattice.
It can be seen that a cross-over between different behaviors occurs for $y \sim 1$,
with a nonlinear dependence of $\s^2$ on $t$ appearing for large $y$.
\label{thvtyscan}}
\end{center}
\end{figure}

\begin{figure}
\begin{center}
\includegraphics[width=4in]{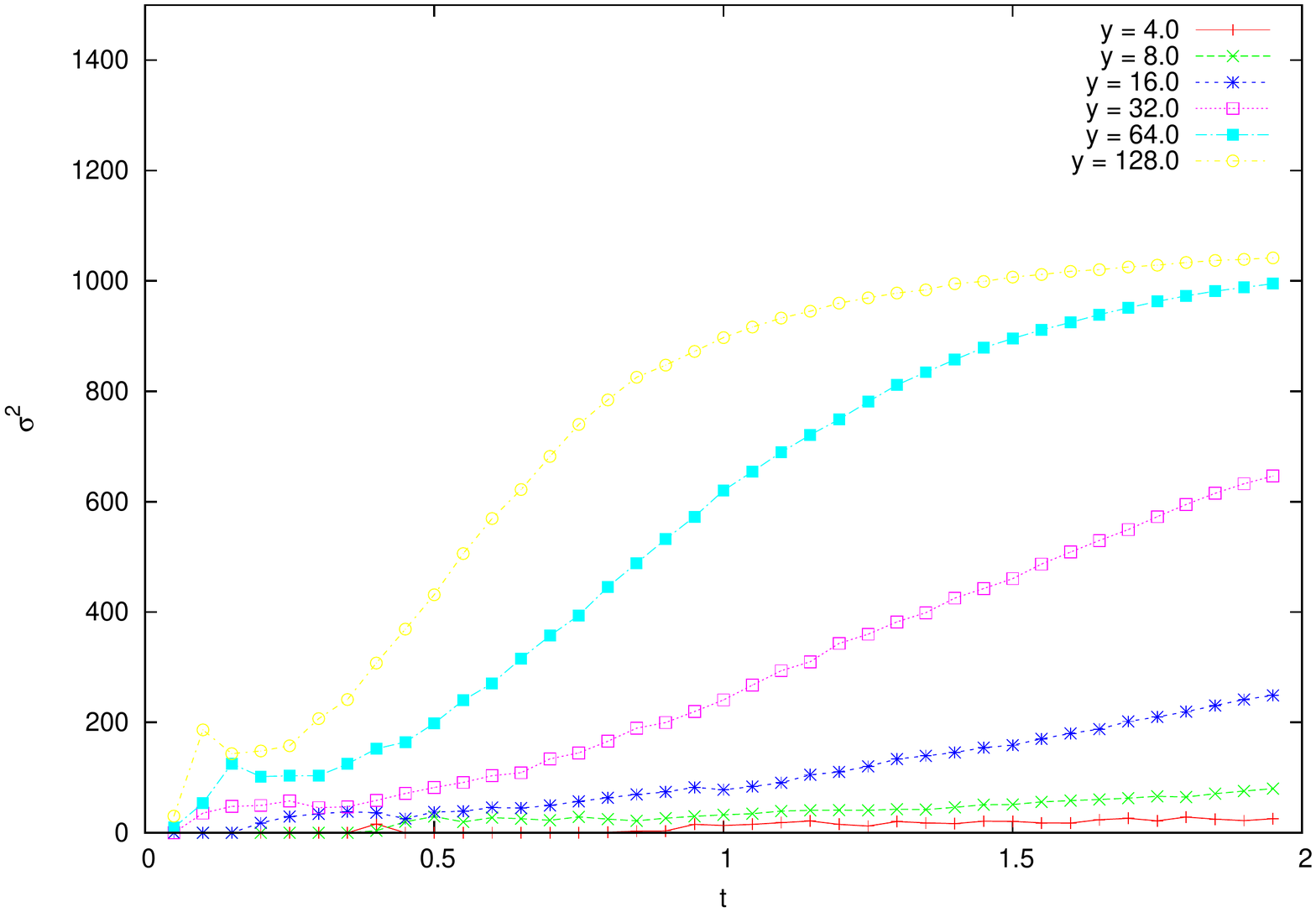}
\caption{Thickness versus $t$, at rather small $t$, for various $y$ for an $L=64$ lattice.
Some interesting features emerge at the smallest values of $t$ seen in this resolution.
It can be seen that the steep rise that was observed in the previous plot simply
becomes steeper, with its center moving to smaller $t$, as $y$ increases.
\label{thvtyscan_tfine}}
\end{center}
\end{figure}

We next turn to results where we hold the fugacity $y = g/2t$ fixed as we vary $t$.    
In Fig.~\ref{thvty010} we show the thickness for $y = 0.1$, which is 
near the IR fixed point where $y=0$.  It can be seen that for these small $t$ values,
and at small $y$, the thickness just behaves as a Gaussian variance directly proportional
to $t$.  It thus appears that the $y$ coupling has essentially no effect in this
regime, other than to determine the slope of the line.  A further ellaboration
of the behavior of the thickness as a function of $t$ and $y$ is given
in Figs.~\ref{thvtyscan} and \ref{thvtyscan_tfine}.

\begin{figure}
\begin{center}
\includegraphics[width=4in]{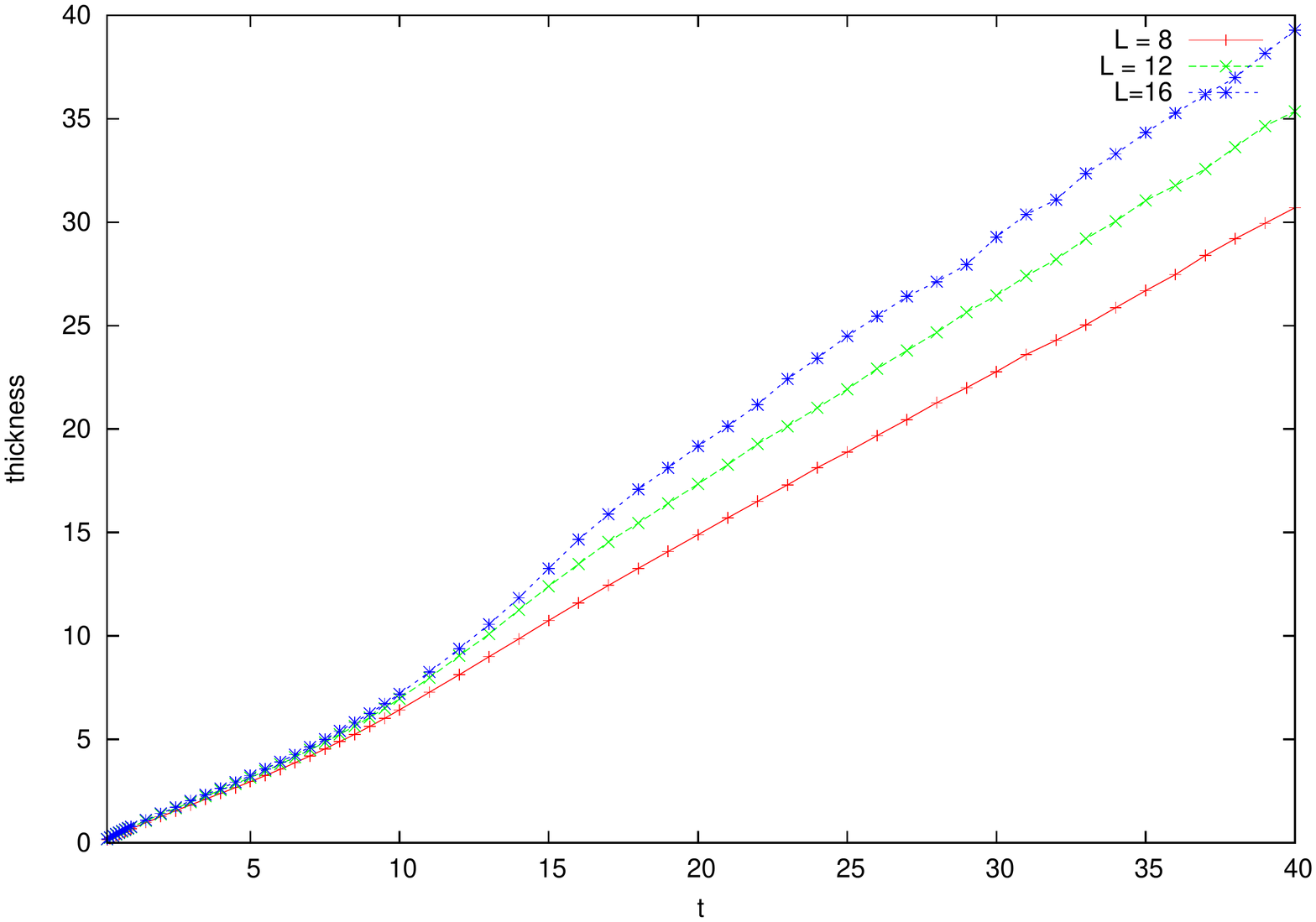}
\caption{The thickness observable rises essentially linearly with $t$ for $y=0.1$.  However,
the slope increases with $L$, as can be seen from these low $L$ values. Note that this
increase in slope only occurs in a noticeable way for $t$ greater than a critical
value of around $10$.  A more precise estimate of $t_c \approx 18$ will be
obtained in our finite size scaling study below, summarized in Fig.~\ref{transition}.  \label{lowLs} }
\end{center}
\end{figure}

In Fig.~\ref{lowLs} we see that the approximately linear behavior $\s^2 \sim t$ has
a slope that rises with $L$, if $t$ is greater than some critical value.
In fact it is this transition that will be used to identify our
critical temperature in the finite size scaling analysis.

\begin{figure}
\begin{center}
\includegraphics[width=4in]{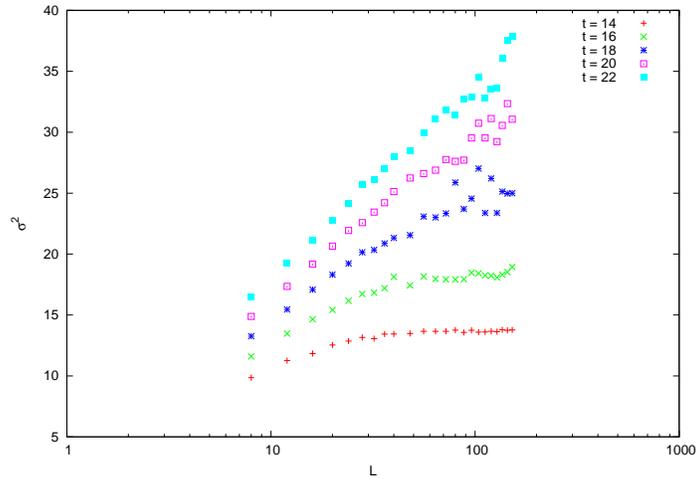}
\caption{The thickness observable transitioning between the
two behaviors \myref{bigLpred} at $t_c \approx 18$.  
Here, we have used $y=0.1$. 
\label{transition} }
\end{center}
\end{figure}

Hasenbusch et al.~provide a perturbative prediction for the finite size scaling
of the thickness above and below the transition temperature \cite{Hasenbusch:1994ef}.
In our conventions it is given by
\beq
\s^2 \simeq \left\{ 
\begin{array}{cc} 
\frac{t}{\pi} \ln L + \text{const.}, & \quad t > t_c \\
\text{const.} & \quad t < t_c
\end{array}
\right.
\label{bigLpred}
\eeq
in the large $L$ limit.  They have also verified this behavior in Monte Carlo
simulations using a cluster algorithm.  We conduct the same study, but with the
Fourier accelerated HMC.  Fig.~\ref{transition} shows precisely this behavior, with $t_c \approx 18$.  We
note that the $y \to 0$ limit (fugacity expansion) predicts $t_c = 8\pi \approx 25$, so there is
apparently some renormalization of $t$ arising from $y = 0.1$.  If we fit the coefficient $c$ of 
$\s^2 \simeq c \ln L + \text{const.}$
for the $t=22$ data, we obtain $c \approx 6.7 \pm 0.2$, which is to be compared with $t/\pi = 7.0$.  
We attribute the small discrepancy ($1.5\s$) to a possible underestimation of errors and
a renormalization of $t$.

There is also an RG interpretation of these results.  The dimension of the cosine
operator in the action is $\Delta = t/4\pi$ (see for instance the discussion
in Section 4.6 of \cite{Fradkin}).  Thus for $t > 8 \pi$, the cosine operator
is irrelevant and the theory flows to a free theory, which will be ungapped
and thus show the logarithmic divergence in the thickness variable.  By contrast
for the $t < 8\pi$ case it will flow to a heavily gapped theory that cuts off
sensitivity to the finite size, and thus gives a fixed result for the thickness
as a function of $L$.

\section{Entropy}
\label{s:ent}
Another important observable is the entropy 
of the system for different values of the fugacity. 
The entropy ${\cal S}$ is calculated using
\beq
{\cal S} = -\( \frac{\p F}{\p t} \)_y, \quad F = -t \ln Z
\eeq
We now show that
\beq
{\cal S}(t) = \ln Z(t) + \vev{ S(t) }
\eeq
where $S(t)$ is the lattice action introduced above,
which has an explicit factor of $1/t$.  
As a first step we therefore write
\beq
S(t) = \frac{1}{t} {\hat S}
\eeq
where now ${\hat S}$ does not depend on $t$.
It is then clear that
\beq
{\cal S} &=& \ln Z + \frac{t}{Z} \frac{\p }{\p t} \int [d\phi] e^{- \frac{1}{t} {\hat S}}
\nnn &=& \ln Z + \frac{1}{Z} \int [d \phi] \frac{1}{t} {\hat S} e^{-S}
\nnn &=& \ln Z + \vev{S}
\eeq

\begin{figure}
\begin{center}
\includegraphics[width=5in]{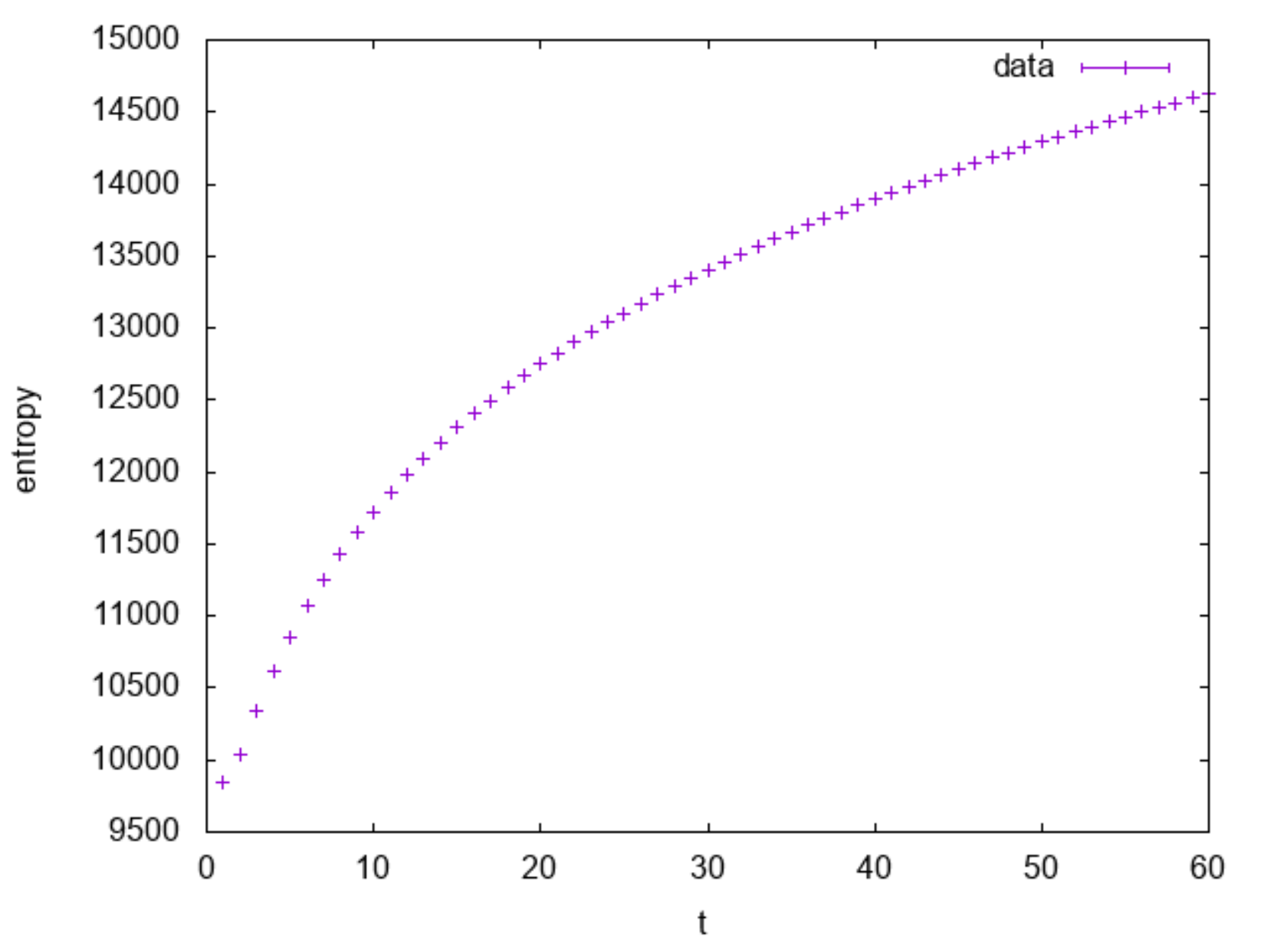}
\caption{Entropy vs.~temperature for fugacity $y=0.1$.
\label{efe_fig1}}
\end{center}
\end{figure}

\begin{figure}
\begin{center}
\includegraphics[width=5in]{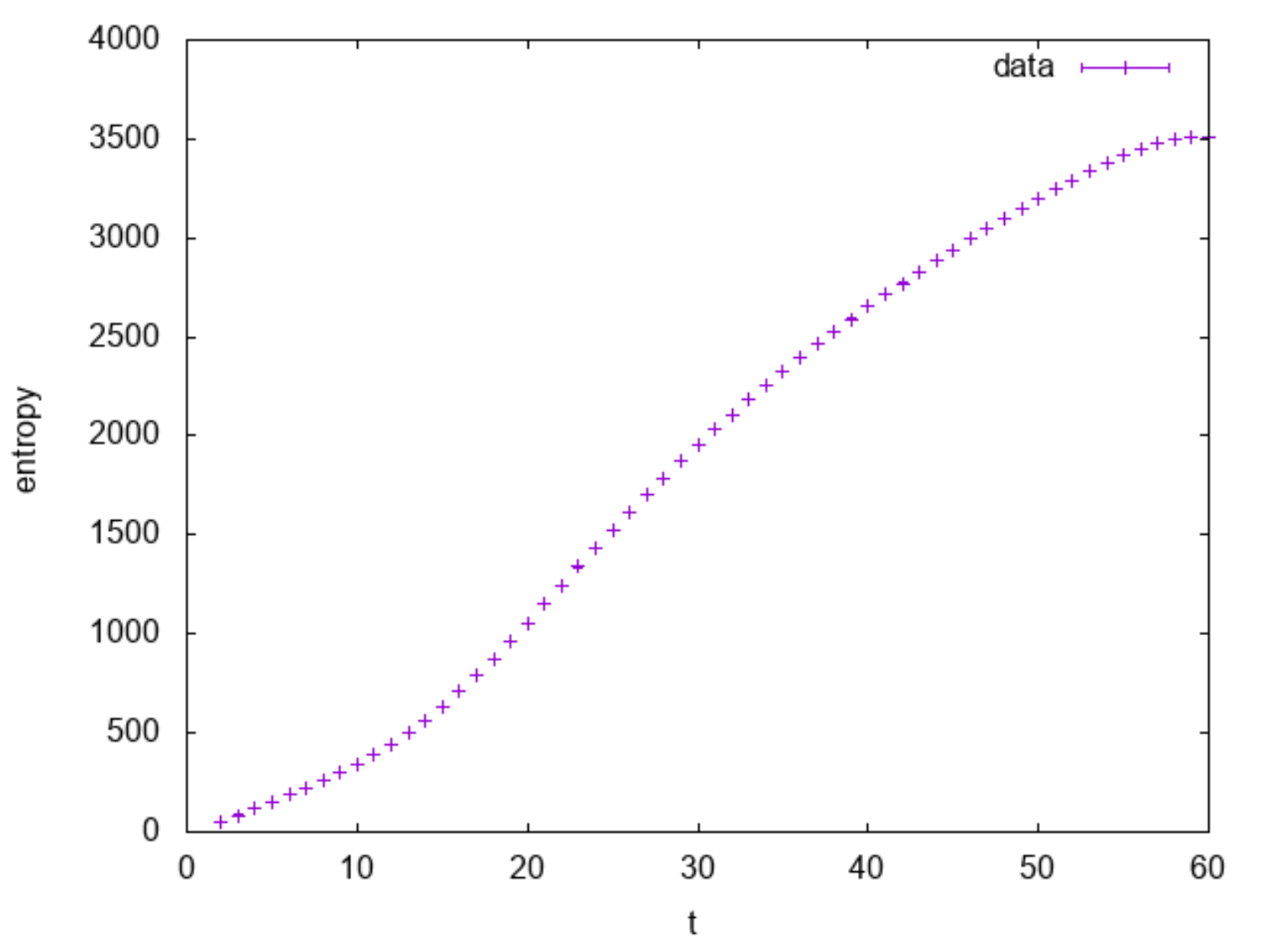}
\caption{Entropy vs.~temperature for fugacity $y=1$.
\label{efe_fig2}}
\end{center}
\end{figure}

\begin{figure}
\begin{center}
\includegraphics[width=5in]{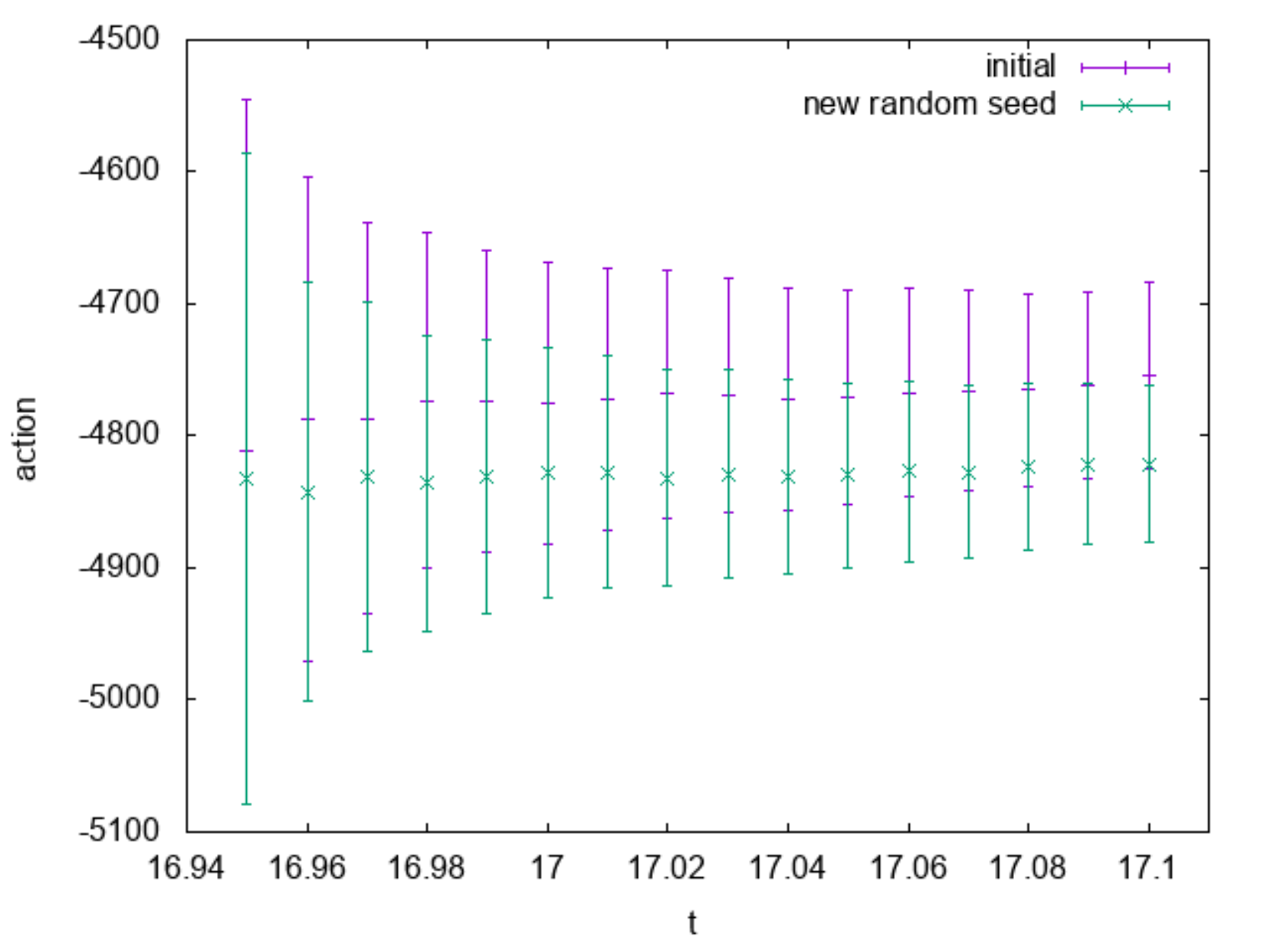}
\caption{
Average action vs.~temperature for fugacity $y=1$.
\label{efe_ac_fig1}
}
\end{center}
\end{figure}

Next we must discuss the computation of $\ln Z(t)$.  With Monte Carlo sampling, we can only do this
relative to some fixed value $Z(t_0)$.  Thus we use
\beq
\int_{t_{0}}^{t} \frac{d}{dt} ~ \ln Z(t) ~ dt 
= \ln \frac{ Z(t) }{ Z(t_{0}) }
\eeq
To find $d \ln Z(t) / dt$ we use the identity
\beq
\frac{d}{dt} \ln Z(t) = \frac{1}{t} \vev{S(t)}
\eeq
from above.  Thus, we can obtain $\ln[Z(t)/Z(t_0)]$ from a straightforward expectation
value---exactly what is easily calculated in the course of a Monte Carlo
simulation.

For a select set of $t$, the action is calculated for 50,000 configurations, 
where the data starts to be recorded after 10,000 steps. 
This allows the system to thermalize. We perform this calculation for a wide range of $t$
and fit $\vev{S(t)}/t$ to a high order polynomial.  The polynomial is then integrated to
obtain $\ln[Z(t)/Z(t_0)]$.  For the reference value $t_0$, we choose $t_0=1$.

An example of the entropy analysis can be seen in Fig.~\ref{efe_fig1} where entropy is 
calculated for a temperature range of $[0,60]$ for fugacity $y=0.1$. 
There appears to be a smooth logarithmic increase in entropy with respect to $t$. 
This makes sense, as a BKT transition is a transition of infinite order: a first derivative
of the free energy is not expected to show a discontinuity, and in fact should be
smooth.  To confirm this expectation, in Fig.~\ref{efe_fig2} another example is shown for a higher 
fugacity, $y=1$.  Due to the higher potential barrier, tunneling was a 
lot more erratic across temperatures, so the action required a higher order polynomial 
to reliably model the free energy.  And as can be seen in the figure, 
the resulting curve is smooth with a monotonically increasing entropy. 

It is important to consider the effects of autocorrelation in this calculation of entropy as well. 
The observable is highly local, since it is a sum over the action density.  The result
is that the autocorrelation is significantly smaller than in the case of
the thickness measurements, which involve correlations over large scales.
In order to quantify this smaller autocorrelation, in Fig.~\ref{efe_ac_fig1} we show the action for $y=1$ 
generated for two separate random seeds, which alters the initialization of $\phi(x)$ at the
beginning of the simulation.  The difference between the action on the two
sets of configurations is well within the error bars, which take into
account integrated autocorrelation time in the usual way.

\section{Clustering}
\label{s:clu}
\begin{figure}
\begin{center}
\includegraphics[width=5in]{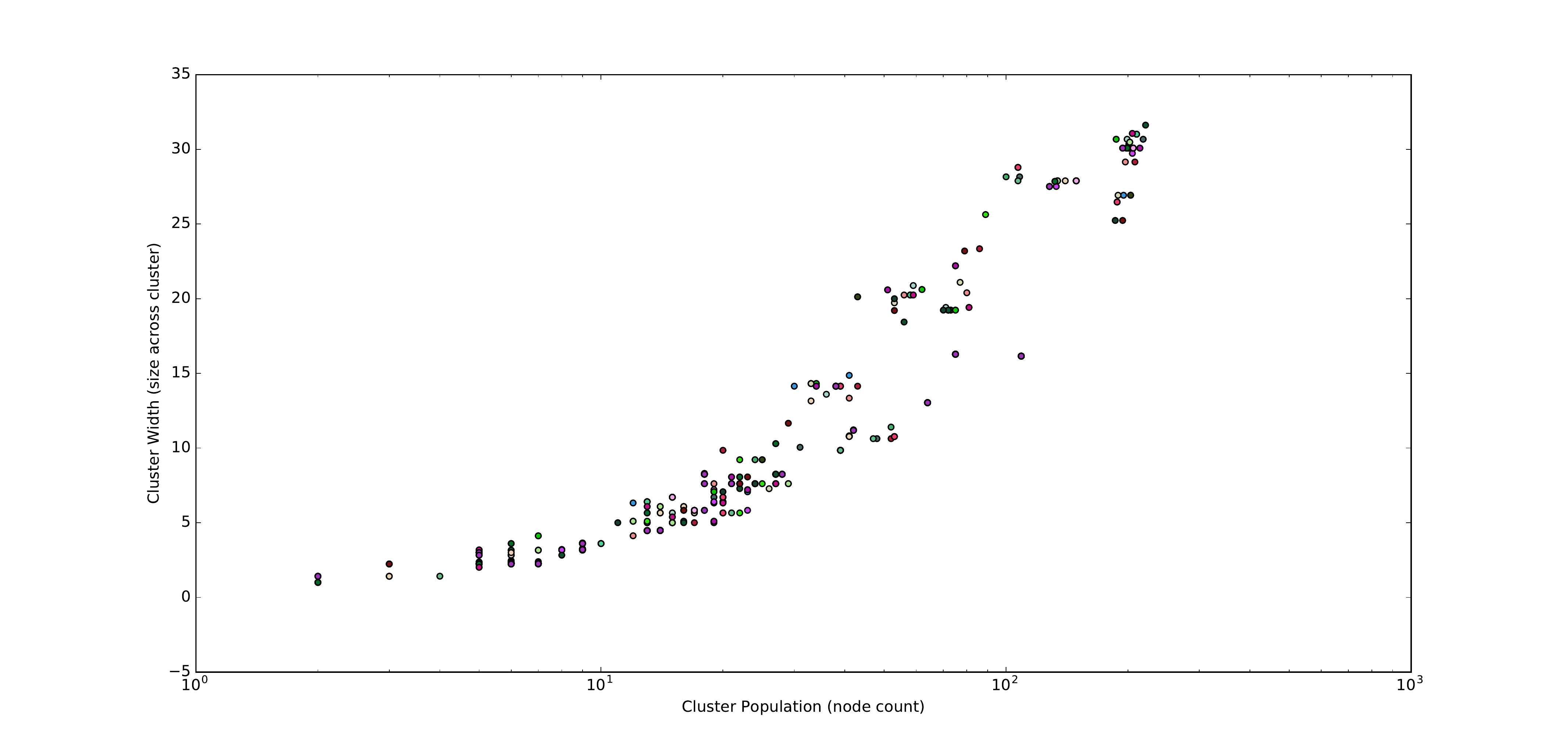}
\caption{Correlation between cluster population and cluster ``width.''  The width of a cluster
is the maximum size across in the two-dimensional domain.  The population is the number of
sites, or ``nodes,'' that have been identified as belonging to the cluster.  Naturally the
wider the cluster, the more nodes that it contains.  However, the precise relationship
is not so easily predicted, and empirically is logarithmic, according to Eq.~\myref{widthreln}.
\label{clustcorr}}
\end{center}
\end{figure}

We have analyzed the clustering of the field $\phi(x)$ into values around $2 n \pi$, $n \in \Zbf$.
The method for identifying a cluster is described in Appendix~\ref{doman}. 
In Fig.~\ref{domain_fig1} the various clusters found by this algorithm are shown from a 
perspective where they are all visible.  Fig.~\ref{domain_fig1} shows the same collection of
clusters edge-on, where it is apparent that they are grouped into layers.  
In order to produce results with clearly defined domains, we found that it was helpful to use a 
random start with $\phi(x)$ drawn uniformly from $[-20,20]$ and a very large fugacity
of $y=15$.  We also set $t=0.1$ for this simulation, which tends to
freeze the configuration into domains (very low temperature).  
The correlation between cluster ``width'' $W$ and population $P$ 
(the number of sites contained in the cluster) is show in Fig.~\ref{clustcorr}, 
where 50 lattice configurations were combined.  
It shows an approximate behavior of 
\beq
W \sim (\ln P)^2
\label{widthreln}
\eeq
Note that the largest population and width is limited by the fact that this is an $L=64$ lattice.
We are not aware of a theoretical argument explaining the behavior \myref{widthreln},
but leave it as a question for future investigations.  

\begin{figure}
\begin{center}
\includegraphics[width=5in]{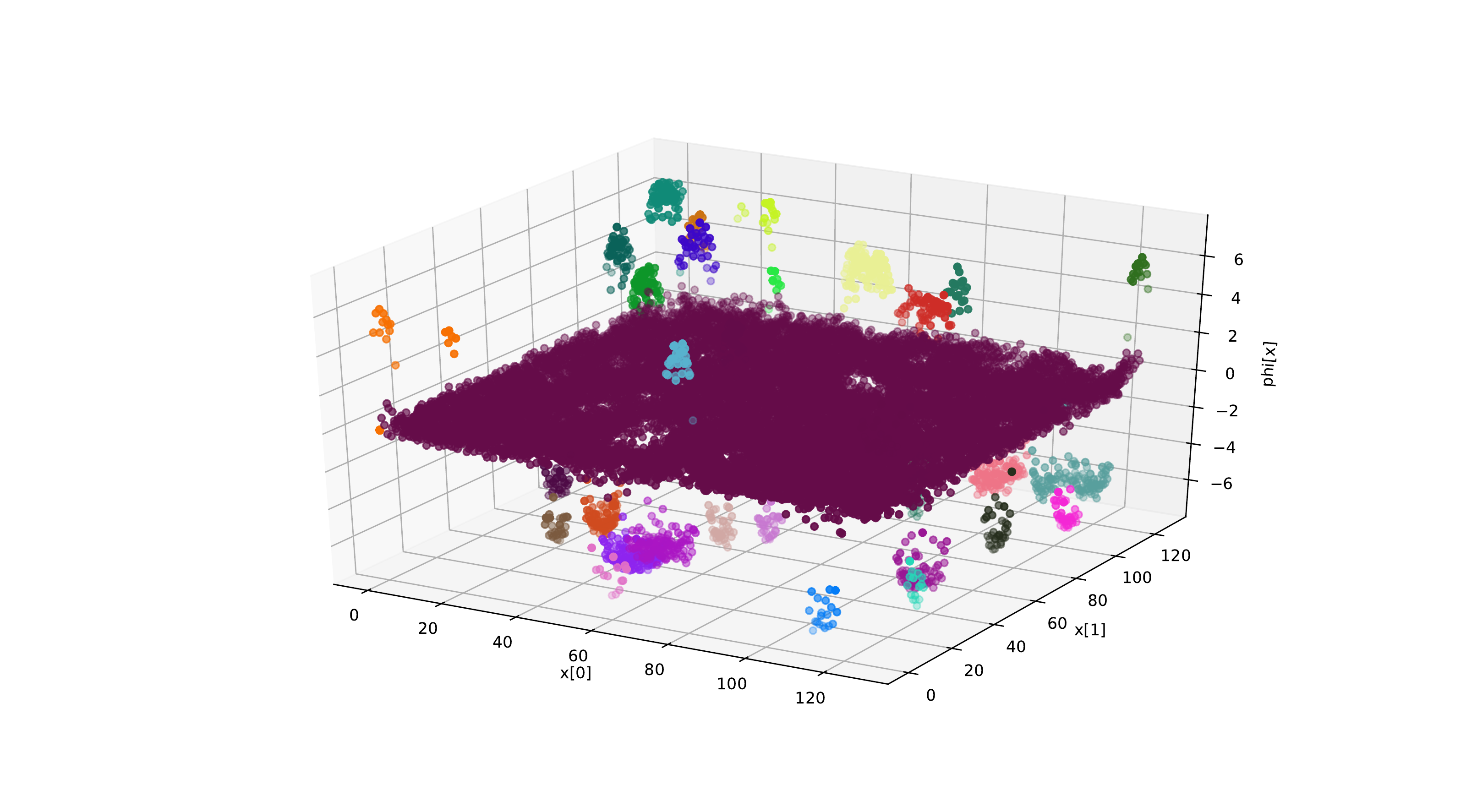}
\caption{The clustering of the field $\phi(x)$
into domains for a particular (thermalized)
configuration with $t=0.1$, $y=15$ and $L=64$.
It can be seen that the bulk of the points cluster
around $\phi=0$, but that other smaller clusters
do form around the vacua at $\phi = \pm 2 \pi$.
\label{domain_fig1}}
\end{center}
\end{figure}

\begin{figure}
\begin{center}
\includegraphics[width=5in]{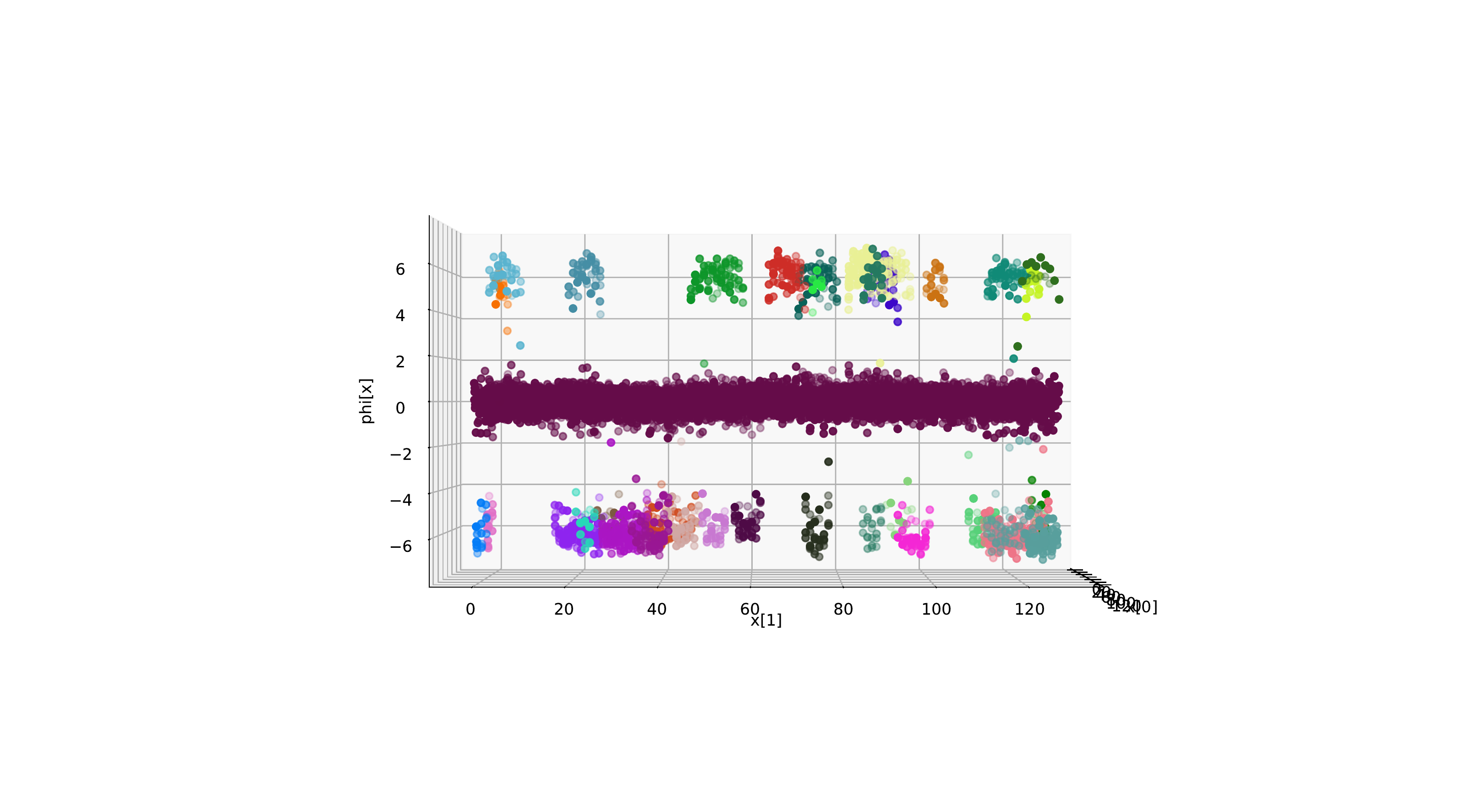}
\caption{The same configuration as in Fig.~\ref{domain_fig1}, but
from a different perspective.
\label{domain_fig2}}
\end{center}
\end{figure}

\begin{figure}
\begin{center}
\begin{tabular}{cc}
\includegraphics[width=3in]{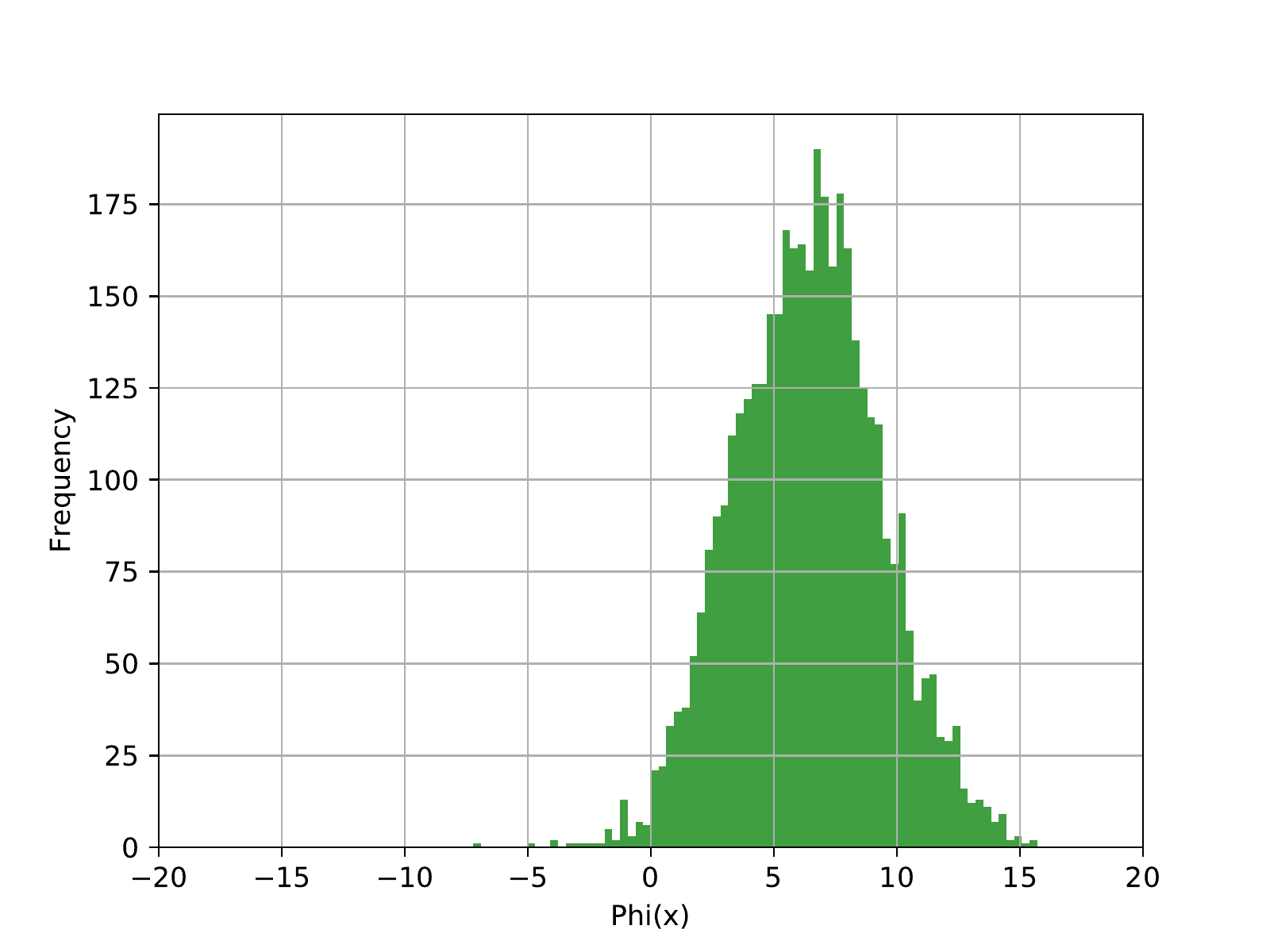} &
\includegraphics[width=3in]{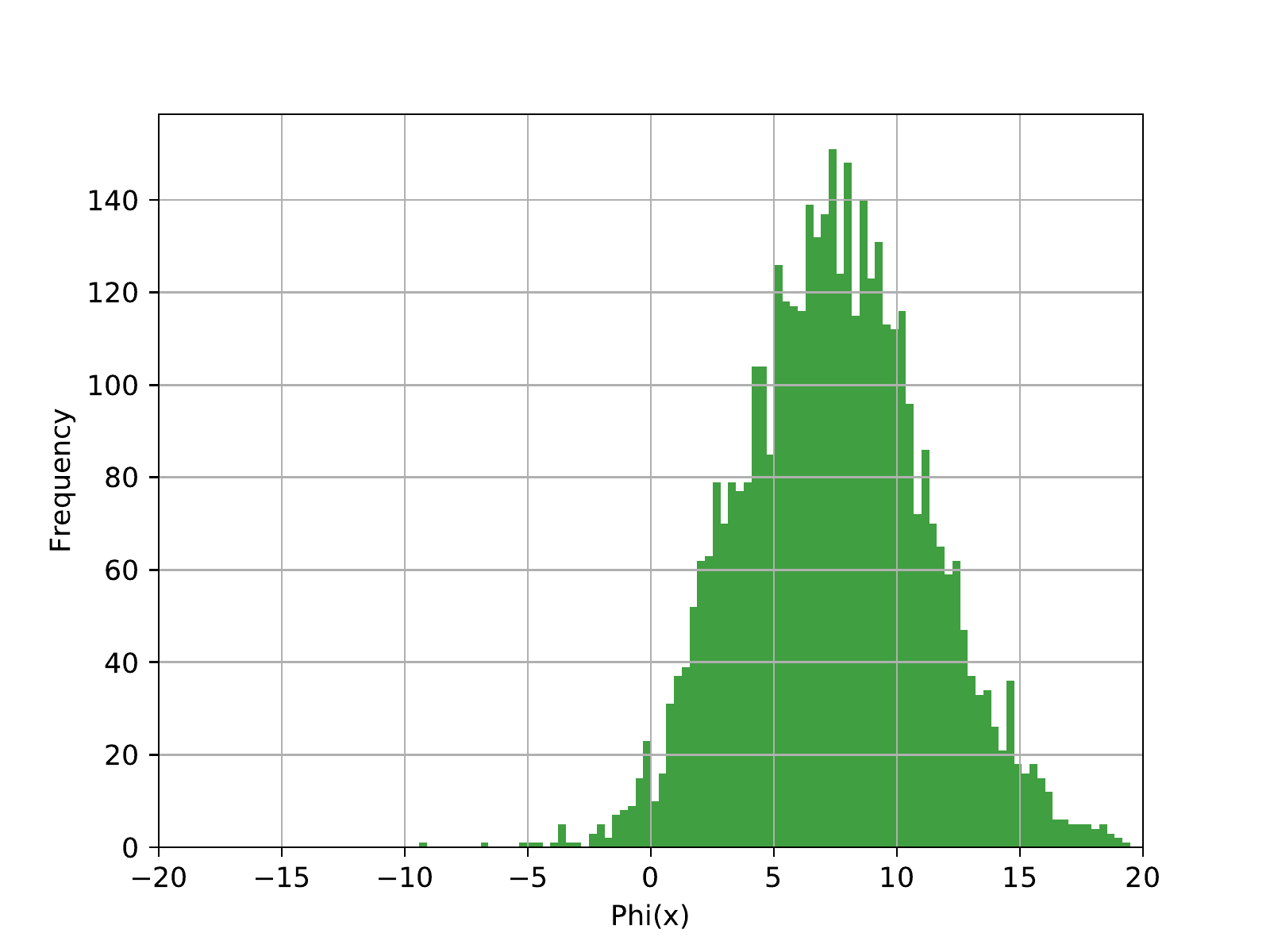} \\
(a) & (b) \\
\includegraphics[width=3in]{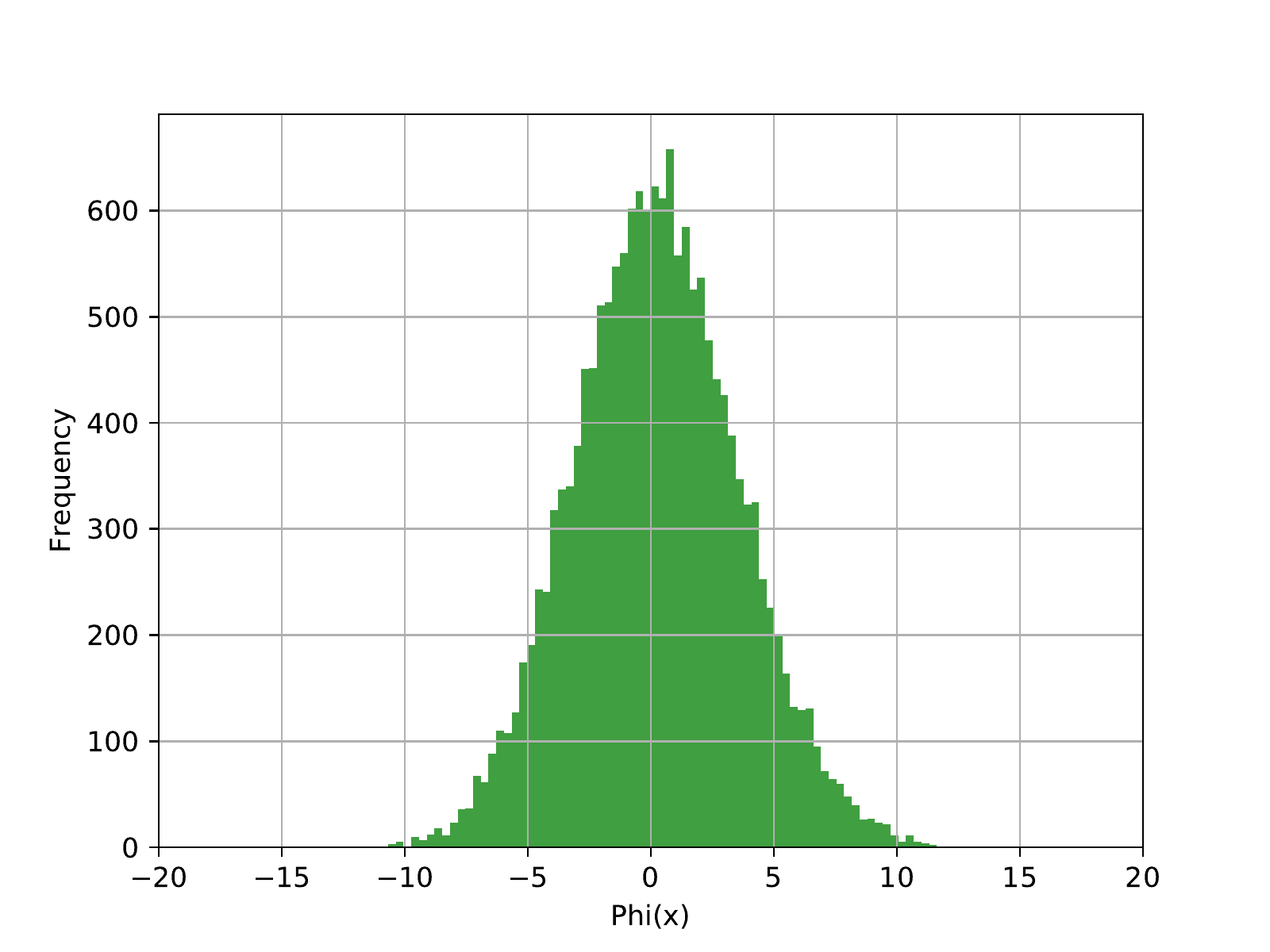} &
\includegraphics[width=3in]{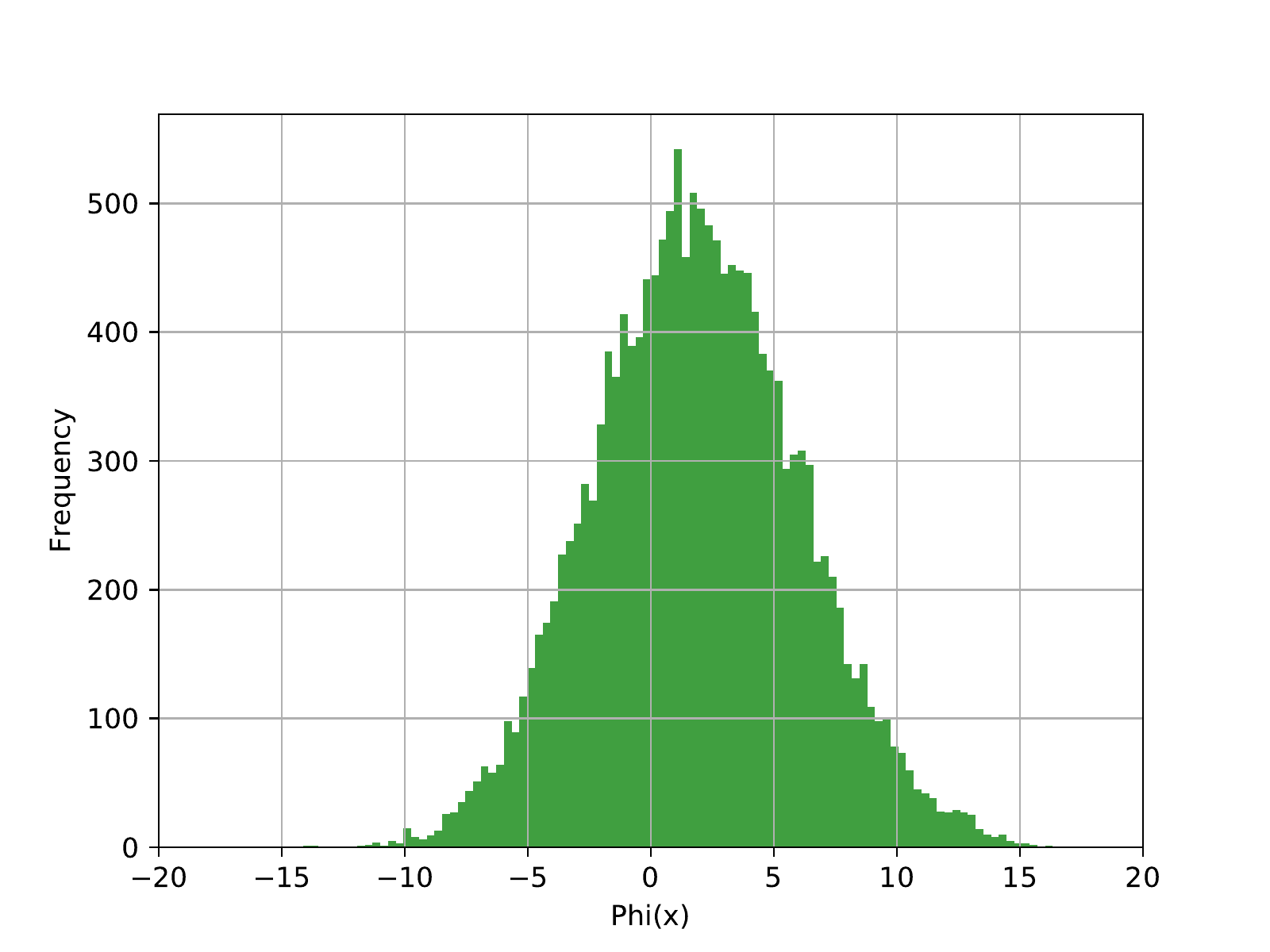} \\
(c) & (d) \\
\end{tabular}
\caption{Distribution of field values on either
side of the transition.  Here, $L=64$ and $y=0.1$.  Panel
(a) is $t=14$ and panel (b) is $t=22$.  These are the
distributions for a single configuration, after
10000 thermalization updates.  The two simulations were
started from the same initial conditions with the same
seed for the pseudorandom number generator, so that
the differences are entirely due to the value of $t$.
It can be seen that $t=22$ has longer tails, due to
stronger fluctuations at the higher temperature.
However, it is interesting that for this $L$, nothing
dramatic occurs as we cross the transition.  In (c) and (d)
we show the same analysis for L=128.  
\label{histo1422}}
\end{center}
\end{figure}

\section{Conclusions}
\begin{figure}
\begin{center}
\includegraphics[width=6in]{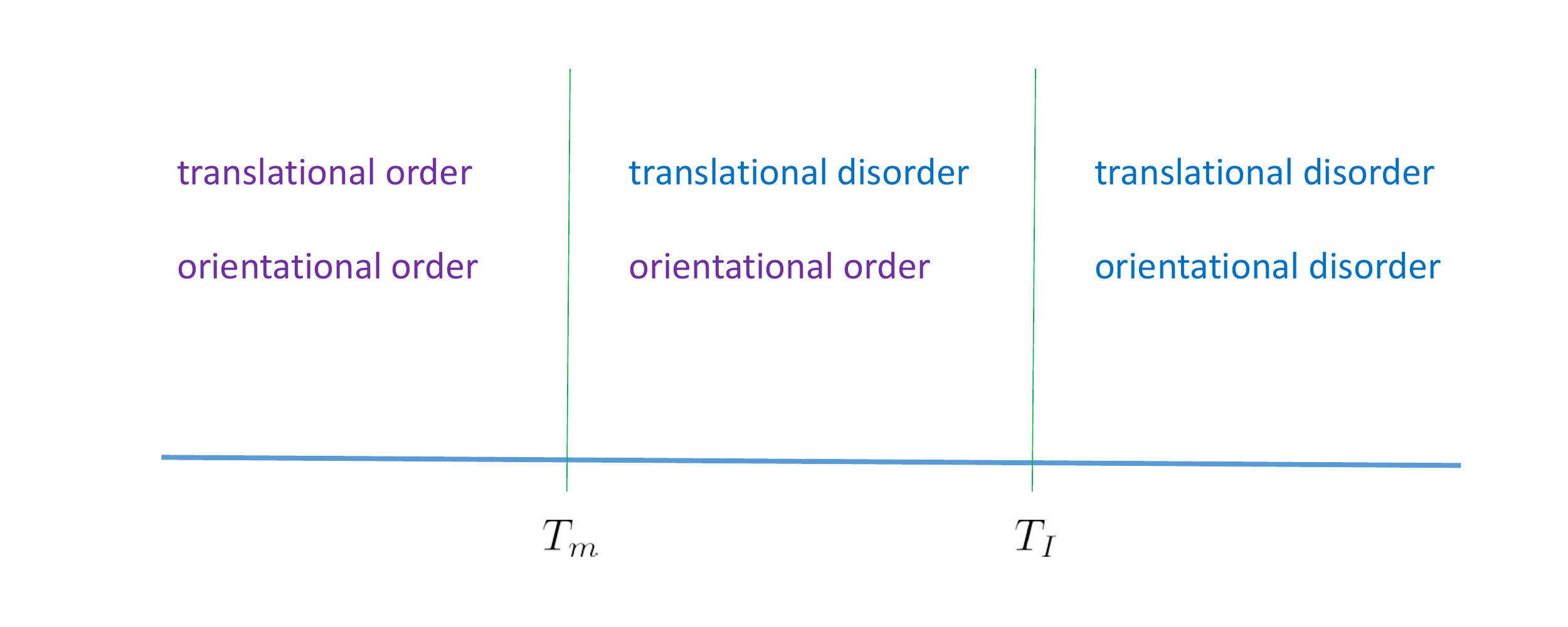}
\caption{The three phases of algebraic order versus disorder, as advanced in the seminal
papers of Halperin, Nelson and Young (1978-79). 
\label{nelson} }
\end{center}
\end{figure}

One interesting avenue to explore is the possible existence of three phases in the XY type
systems, as shown in Fig.~\ref{nelson}.  Arguments for the existence of this more complicated
set of transitions were advanced in classic papers by Halperin, Nelson and Young 
\cite{Halp78a,Halp79a,Young79,Nelson78}.  Although this feature has been argued for
theoretically, and observed experimentally, it has still not been detected in
numerical simulations to date.  Thus, future work would be to use the more powerful
and sophisticated simulation methods that are begun here to observe these
types of behaviors.  An important question is:  ``How would they be realized
in the sine-Gordon model, if at all.''  Does a theory of vortices contain enough
information to capture the full dynamics of the XY model in this regard?
We leave this question open to future research.

We mentioned in the main body of the paper that in superconducting
thin films one can understand the behavior as a superposition of BKT
fluctuations (spin waves and vortices) and GL fluctuations.
It would seem that the presence of GL fluctuations comes from the fact that
the thin film is not truly two-dimensional, having a thickness $d$.
This suggests an XY model with three dimensions, one of them being quite
small.  Alternatively, we could formulate a SG model built
on layers.  One would like to establish the connection between
these ideas.  Currently we are developing numerical simulations
that will analyze such layered systems, especially in the
limit of weak interlayer interactions.  

\section*{Acknowledgements}
JG was supported in part by the Department of Energy, Office of Science, Office of High Energy Physics,
Grant No. DE-SC0013496. 
Significant parts of this research were done using resources provided by the Open Science Grid 
\cite{osg1,osg2}, 
which is supported by the National Science Foundation award 1148698, and the U.S. Department of Energy's Office of Science.
We are also appreciative of XSEDE \cite{xsede} resources (Stampede),
where other significant computations for this study were performed.

\appendix

\section{Free theory propagator}
\label{ftprop}
\subsection{Continuum}
We generalize the free action to include a mass term,
\beq
S_0(\theta) = \frac{1}{t} \int d^2 x ~ \bigg\{ \half [ \p_\mu \phi(x) ]^2 + \half m^2 \phi^2(x) \bigg\}
\eeq
We must include a mass term to regulate the infrared divergence.  Then
\beq
\Delta(x,m) = \int \frac{d^2p}{(2\pi)^2} \frac{e^{i p \cdot x}}{p^2 + m^2}
\eeq
This can be evaluated with the help of results from Gradshteyn \& Ryzhik (G\&R) \cite{GR94}.
Passing to polar coordinates and using G\&R 3.339 (p.~357)
\beq
\frac{1}{4 \pi^2} \int_0^\infty \frac{p dp}{p^2 + m^2} \int_0^{2\pi} d\theta ~ e^{ipx\cos\theta}
= \frac{1}{2\pi} \int_0^\infty \frac{p dp}{p^2 + m^2} J_0(px)
\eeq
Then using G\&R 6.532\#4 (p.~702) we perform the integral obtaining
\beq
\frac{1}{2\pi} K_0(mx)
\eeq
We are interested in the limit of small $m$, which is obtained from G\&R 8.447 \#3 \& \#1 (p.~971)
\beq
K_0(z) = -\ln \frac{z}{2} + \psi(1) + \ord{z}, \quad \psi(1) = - \gamma
\eeq
Thus we finally obtain
\beq
\Delta(x,m) = -\frac{1}{2\pi} \ln \frac{mx}{2} - \frac{1}{2\pi} \gamma + \ord{m}
= -\frac{1}{4\pi} \[ \ln \frac{m^2 x^2}{4} + 2 \gamma \] + \ord{m}
\label{Delta_eq}
\eeq
which agrees with ZJ (32.6) \cite{ZinnJustin}.

Next we consider the correlation functions of composite operators as appears in \myref{zeroth},
making use of \myref{Delta_eq} to prove this identity.  The trick (e.g.~\cite{ZinnJustin}) is to compute the path
integral
\beq
\left\langle \prod_{i=1}^n e^{i \e_i \phi(x_i)} \right\rangle =
Z_0^{-1} \int [d\phi(x)] e^{-S_0[\phi(x)]} \prod_{i=1}^n e^{i \e_i \phi(x_i)}, \quad
Z_0 = \int [d\phi(x)] e^{-S_0[\phi(x)]}
\eeq
by rewriting it in terms of a ``source''
\beq
J(x) = i \sum_{i=1}^n \e_i \delta(x - x_i)
\label{myJcur}
\eeq
so that
\beq
&& \left\langle \prod_{i=1}^n e^{i \e_i \phi(x_i)} \right\rangle =
\ddd Z_0^{-1} \int [d\phi(x)] \exp \left( - \int d^2x ~ \left\{ \frac{1}{2t} (\p_\mu \phi(x))^2 
+ \frac{m^2}{2t} \phi^2(x) - J(x) \phi(x) \right\} \right)
\eeq
This can be computed, as usual, by completing the square, leading to
\beq
\left\langle \prod_{i=1}^n e^{i \e_i \phi(x_i)} \right\rangle =
\exp \left( \frac{t}{2} \int d^2x ~ d^2y ~ J(x) \Delta(x-y;m) J(y) \right)
\eeq
Substituting \myref{myJcur}
we obtain
\beq
\left\langle \prod_{i=1}^n e^{i \e_i \phi(x_i)} \right\rangle =
 \prod_{i,j=1}^n \( \frac{m^2 (x_i-x_j)^2}{4} \)^{\frac{t}{8\pi} \e_i \e_j}
\eeq
To get closer to 
\myref{zeroth}
we separate out factors
\beq
\lim_{\delta\to 0} \prod_i \( \frac{m \delta}{2} \)^{\frac{t}{4\pi} \e_i^2} \cdot
\prod_{i<j} \( \frac{m |x_i-x_j|}{2} \)^{\frac{t}{2\pi} \e_i \e_j}
\label{cofactored}
\eeq
where various factors of 2 have been accounted for in the exponents; in particular
the double-counting of pairs in $\prod_{ij}$ versus $\prod_{i<j}$.  The $\delta\to0$
limit occurs because of the coincident ``diagonal'' terms $x_i=x_j$ in $\prod_{ij}$,
and will in practice be regulated by lattice expressions for the propagator. 
I.e., they correspond to
\beq
\lim_{x_i \to x_j} \int_{-\pi}^\pi
\frac{d^2 k}{(2\pi)^2} 
\frac{e^{i k \cdot (x_i - x_j) }}{ {\hat k}^2 + m^2 }
\eeq
where ${\hat k}_\mu = 2 \sin ( k_\mu / 2 )$.  Because of the compact domain
of integration (finite lattice spacing), there is no singularity associated
$x_i \to x_j$.  The second factor of \myref{cofactored} agrees with \myref{zeroth}, in terms
of the dependence on $|x_i-x_j|$; everything else is swept into the renormalization
factor $\zeta$, while introducing the renormalization scale $\mu$.

\section{Physics of vortices}
\begin{figure}
\begin{center}
\includegraphics[width=3in]{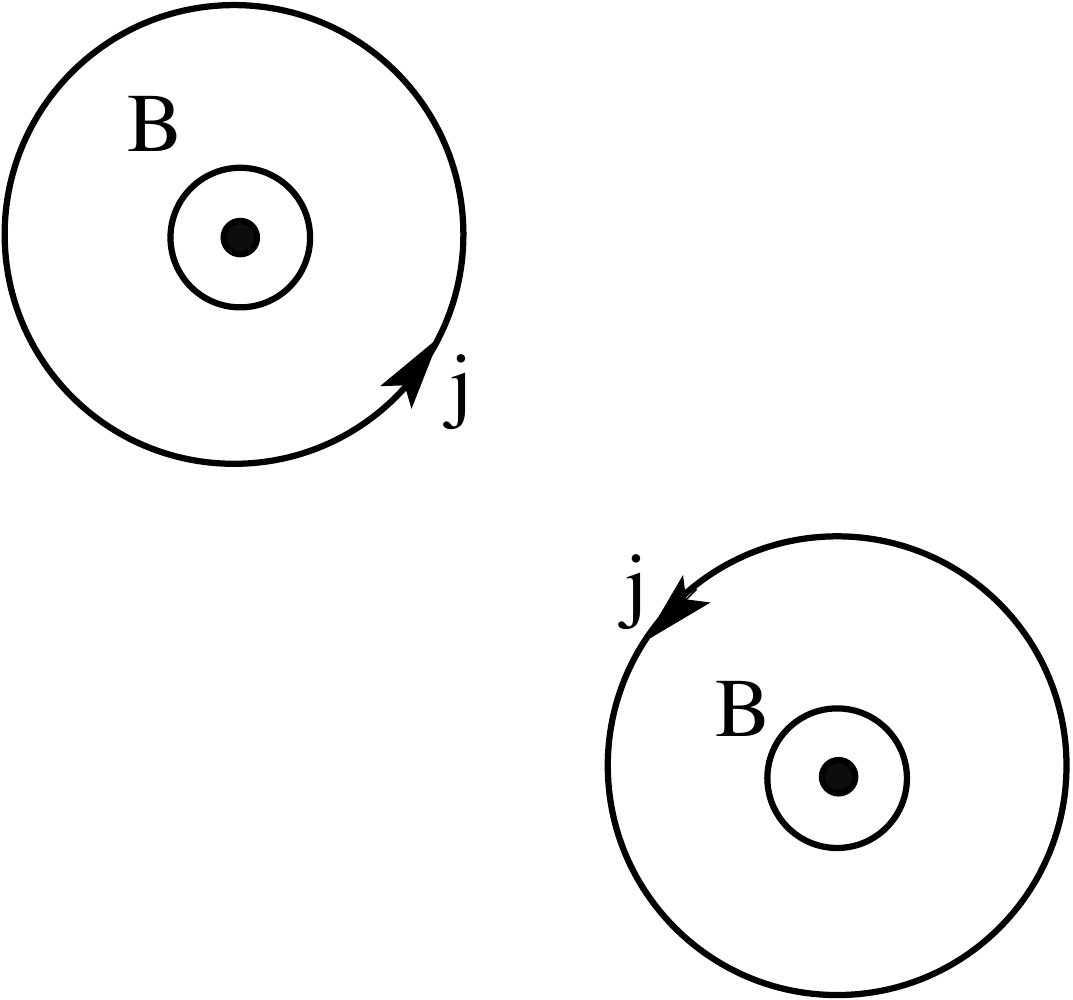}
\caption{Two vortices and their eddy currents. \label{vort3}}
\end{center}
\end{figure}

The sine-Gordon model is supposed to capture the physics of vortices, where the field
variable $\phi(x)$ represents the angle of the $U(1) \simeq O(2)$ phase of the
order parameter:  for instance the condensate in a thin film superconductor, or the
phase of the wavefunction in a superfluid.  In
a superconductor the vortices describe how magnetic field lines penetrate the bulk,
and surrounding each vortex is an eddy current.  If one considers two vortices
such as in Fig.~\ref{vort3}, one can see that the eddy currents will tend to repel since the portions
of the currents that are closest are going in opposite directions.\footnote{Recall that
unlike charges, for currents opposites repel.}

\section{Map between different forms of the action}
\label{s:maps}
An alternate form of the action that we have used early in our simulations is
\beq
S'[\phi] = \frac{1}{t'} \int d^2 x ~ \left\{ \half [ \p_\mu \phi'(x) ]^2 - g' \cos( 2 \phi'(x) ) \right\}
\label{sgactp}
\eeq
This can be related to the action in the body of the paper, Eq.~\myref{sgact},
by defining $\phi(x) = 2 \phi'(x)$ and relating
$(t,g)$ to $(t',g')$ appropriately:
\beq
t = 4t', \quad g = 4 g'
\eeq

In \cite{Hasenbusch:1994ef} the following convention is used:
\beq
S[\varphi] = \int d^2 x ~ \left\{ \frac{1}{2\beta} (\p_\mu \varphi)^2 - z \cos (2 \pi \varphi) \right\}
\eeq
The map is then
\beq
\varphi = \frac{1}{2\pi} \phi, \quad \beta = \frac{t}{4 \pi^2}, \quad z = \frac{g}{t}
\eeq
The thicknesses are related by
\beq
\s_H^2 = \frac{1}{4\pi^2} \s_J^2
\eeq
where on the l.h.s.~there is the Hasenbusch et al.~convention, and on the r.h.s.~is ours.
As a result we end up with the prediction \myref{bigLpred} from their result
\beq
\s_H^2 \simeq \frac{\beta}{\pi} \ln L + \text{const.}
\eeq

\section{Domain analysis}
\label{doman}
For every configuration, the values of $\phi$ tend to group together into distinct domains during the thermalization of the system. Plotting the system in terms of x, y, and $\phi$ yields a $L \times L$ scatter plot of particles as seen in Fig.~\ref{domain_fig2}. For this particular configuration the potential barrier is low enough, and the temperature is high enough that sites are able to tunnel to several domains above and below $\phi=0$, known as the threshold.

The combination of fugacity and temperature for used to produce the
configuration plotted in Fig.~\ref{domain_fig2} makes it easy for an algorithm to recognize
the clusters.  In particular, it allows K-means, a prototype-based cluster algorithm, to efficiently
identify the clusters and their centroids. One major issue that is present in using K-means, however, 
is that this algorithm requires a predefined cluster count, which is not easily recognizable for large 
lattice sizes; and due to the nature of the field-versus-site scattering, the commonly-used elbow method is not 
inherently helpful. We therefore used an alternative: density-based cluster definitions. Here the cluster count 
is defined using the DBSCAN method, following the format of using the $k-dist$ values to define cluster radius and 
minimum point threshold. Once $k$ number of cluster radii (called $Eps$) are identified, the algorithm 
uses K-means to classify those clusters. To aid in speeding up the process, centroids $c_{1}...c_{k}$ are 
initialized within the local space of each $Eps_{j}$, automatically assigning identified core points 
and border points to respective cluster $c_{j}$. For any points missed, we simply follow ${\rm argmin} \, D(x_{i},c_{j})$, 
where $D$ is distance, assigning the noise point $x_{i}$ to closest cluster $c_{j}$.

\begin{figure}
\begin{center}
\includegraphics[width=5in]{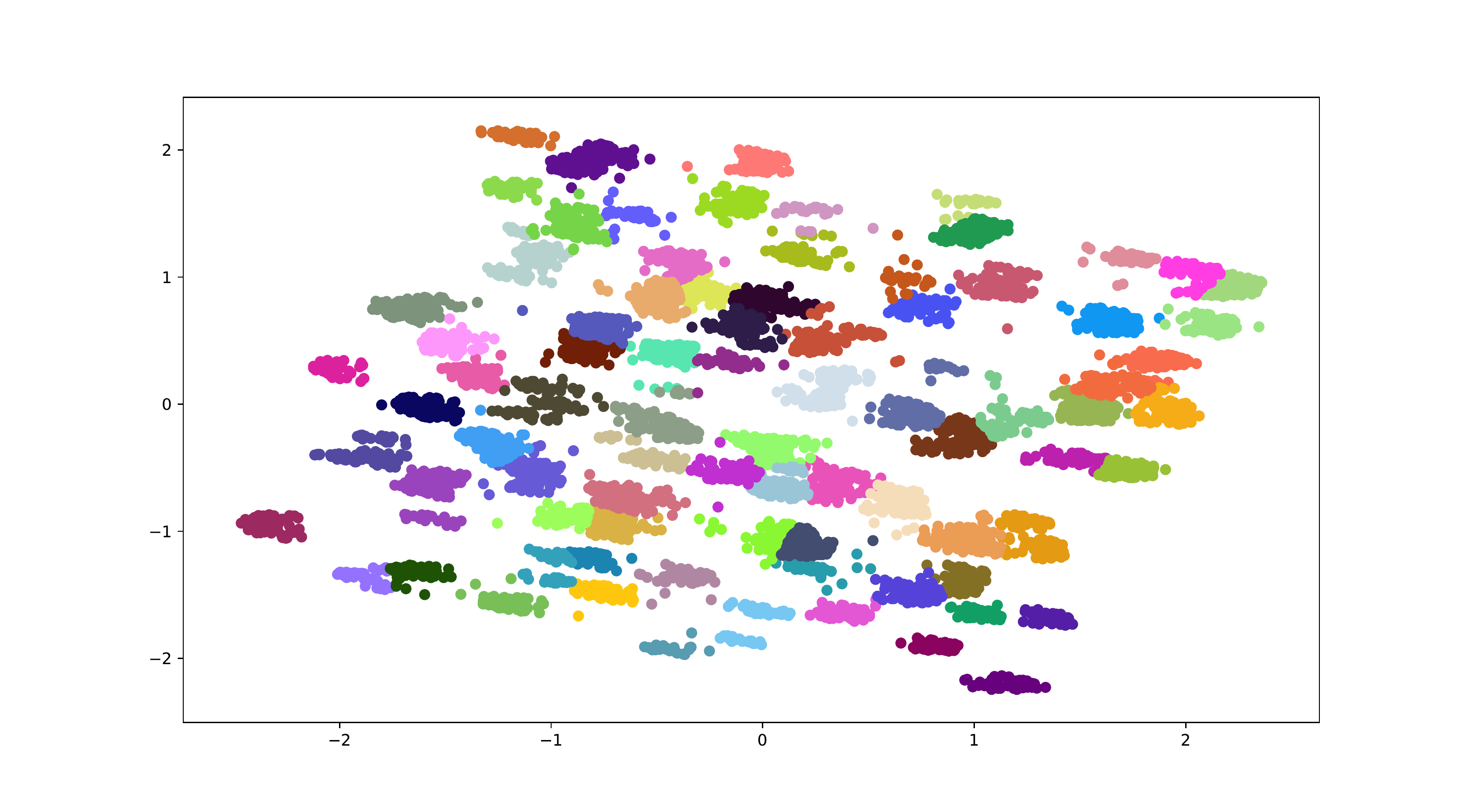}
\caption{PCA of (thermalized) configuration 1000 for $L=256$, $y=4$, $t=1$.
\label{pca_fig1}}
\end{center}
\end{figure}

The algorithm is quick and efficient, but is still not powerful enough when applied to even 
larger lattice sizes coupled with small deviations of $\phi$. Many configurations of this form
can be too complex for this algorithm. A work-around for this is to analyze extremely large amounts
of sites is to take the configuration data and project it onto a 2d plane using principle component analysis (PCA). 
In PCA, a system with a certain amount of unlabeled data is selected to form a $s$-dimensional sample space. 
The covariance matrix is then computed as 
\beq
\Sigma = 
 \begin{pmatrix}
  \sigma_{1,1} & \sigma_{1,2} & \cdots & \sigma_{1,j} \\
  \sigma_{2,1} & \sigma_{2,2} & \cdots & \sigma_{2,j} \\
  \vdots  & \vdots  & \ddots & \vdots  \\
  \sigma_{i,1} & \sigma_{i,2} & \cdots & \sigma_{i,j} 
 \end{pmatrix}
\eeq
Where $\sigma_{i,i}$ is the variance and $\sigma_{i,j}$ is the covariance. The eigenvalues and eigenvectors are calculated from $\Sigma$, giving a set of potential axes (eigenvectors) that the data could be projected upon. The $j$ eigenvectors with the greatest eigenvalues are chosen and combined with the the $s$-dimensional sample space to form a $s \times j$-dimensional matrix $D$, the reduced subspace. The original samples $X$ are then transformed onto the new subspace $Y$
\beq
Y = D^{T} \times X
\eeq
In terms of our data, we considered the sample space of the configurations as a fibre bundle instead of just using the singular $\phi$ data generated by the simulation. A fibre bundle is, for our simple theory, the product of the base space and the target space. Considering the analysis in terms of a fibre bundle allows us to observe our PCA sample space with three features (x, y, and $\phi$) instead of just $\phi$. This gives depth to our analysis, allowing for the cluster algorithm to identify the clusters quickly for a large number of increasingly large lattice sizes without any loss in accuracy. Fig.~\ref{pca_fig1} shows the PCA of a configuration with a lattice size of 256. Computation time is cut down by
a factor of two and the cluster count and cluster width is not compromised.

\section{Hard core repulsion}
It is straightforward enough to implement the hard-core repulsion in the Coulomb gas model.
For instance the fugacity expansion of the partition function as presented in the main text is\footnote{Recall
that only even powers of the fugacity appear because of the charge neutrality condition.}
\beq
Z_{\text{CG}} = \sum_{n \in 2\Zbf} \frac{y^n}{n!} \int d^2x_1 \cdots d^2x_n
\sum_{\{\e\}} \prod_{i<j} ( \mu |x_i - x_j| )^{\e_i \e_j t/2\pi}
\eeq
where $y=g/2t$.  Hard-core repulsion replaces this with
\beq
Z_{\text{CG-h.c.r.}} = \sum_{n \in 2\Zbf} \frac{y^n}{n!} \int d^2x_1 \cdots d^2x_n
\sum_{\{\e\}} \prod_{i<j} \theta(|x_i - x_j| - R) ( \mu |x_i - x_j| )^{\e_i \e_j t/2\pi}
\eeq
so that the integrand vanishes for $|x_i - x_j| < R$ for any pair $i,j$.  Now the question
is, how would this change the continuum theory---i.e., what becomes of the sine-Gordon model
that this theory is dual to in the absence of the hard-core repulsion?  One path to formulating
an answer is to represent the Heaviside unit step function as
\beq
\theta(x) = \int \frac{d \omega}{2 \pi} \frac{-i e^{i \omega x}}{\omega - i \e}
\eeq
where as usual $\e > 0$ is an infinitesmal number that provides a pole prescription.
Thus we obtain the expression
\beq
Z_{\text{CG-h.c.r.}} &=& \sum_{n \in 2\Zbf} \frac{y^n}{n!} \int d^2x_1 \cdots d^2x_n
\sum'_{\{\e\}} \prod_{i<j} \int \frac{d \omega_{ij}}{2 \pi} \frac{-i e^{i \omega_{ij} |x_i-x_j|}}{\omega_{ij} - i \e}
( \mu |x_i - x_j| )^{\e_i \e_j t/2\pi} \nnn
&=& \sum_{n \in 2\Zbf} \frac{y^n}{n!} \int d^2x_1 \cdots d^2x_n
\sum'_{\{\e\}} \prod_{i<j} \int \frac{d \omega_{ij}}{2 \pi} \frac{-i }{\omega_{ij} - i \e}
e^{ i \omega_{ij} |x_i-x_j| + (\e_i \e_j t/2\pi) \ln |x_i - x_j| }
\eeq
If we define
\beq
\omega_{ij} = (\e_i \e_j t/2\pi) \omega'_{ij}
\eeq
then the expression becomes more amenable to intepretation:
\beq
Z_{\text{CG-h.c.r.}} &=& \sum_{n \in 2\Zbf} \frac{y^n}{n!} \int d^2x_1 \cdots d^2x_n
\sum'_{\{\e\}} \prod_{i<j} \int \frac{d \omega_{ij}}{2 \pi} \frac{-i }{\omega_{ij} - i \e}
e^{ (\e_i \e_j t/2\pi)(i \omega'_{ij} |x_i-x_j| +  \ln |x_i - x_j| }
\eeq
Thus we posit a differential operator $D(\omega)$ that has Green's function
\beq
G(\omega;x,y) = i \omega |x_i-x_j| +  \ln |x_i - x_j|
\eeq
Obviously it is some generalization of the 2d Laplacian.

\bibliography{sinegordon}

\begin{thebibliography}{28}%
\makeatletter
\providecommand \@ifxundefined [1]{%
 \@ifx{#1\undefined}
}%
\providecommand \@ifnum [1]{%
 \ifnum #1\expandafter \@firstoftwo
 \else \expandafter \@secondoftwo
 \fi
}%
\providecommand \@ifx [1]{%
 \ifx #1\expandafter \@firstoftwo
 \else \expandafter \@secondoftwo
 \fi
}%
\providecommand \natexlab [1]{#1}%
\providecommand \enquote  [1]{``#1''}%
\providecommand \bibnamefont  [1]{#1}%
\providecommand \bibfnamefont [1]{#1}%
\providecommand \citenamefont [1]{#1}%
\providecommand \href@noop [0]{\@secondoftwo}%
\providecommand \href [0]{\begingroup \@sanitize@url \@href}%
\providecommand \@href[1]{\@@startlink{#1}\@@href}%
\providecommand \@@href[1]{\endgroup#1\@@endlink}%
\providecommand \@sanitize@url [0]{\catcode `\\12\catcode `\$12\catcode
  `\&12\catcode `\#12\catcode `\^12\catcode `\_12\catcode `\%12\relax}%
\providecommand \@@startlink[1]{}%
\providecommand \@@endlink[0]{}%
\providecommand \url  [0]{\begingroup\@sanitize@url \@url }%
\providecommand \@url [1]{\endgroup\@href {#1}{\urlprefix }}%
\providecommand \urlprefix  [0]{URL }%
\providecommand \Eprint [0]{\href }%
\providecommand \doibase [0]{http://dx.doi.org/}%
\providecommand \selectlanguage [0]{\@gobble}%
\providecommand \bibinfo  [0]{\@secondoftwo}%
\providecommand \bibfield  [0]{\@secondoftwo}%
\providecommand \translation [1]{[#1]}%
\providecommand \BibitemOpen [0]{}%
\providecommand \bibitemStop [0]{}%
\providecommand \bibitemNoStop [0]{.\EOS\space}%
\providecommand \EOS [0]{\spacefactor3000\relax}%
\providecommand \BibitemShut  [1]{\csname bibitem#1\endcsname}%
\let\auto@bib@innerbib\@empty
\bibitem [{\citenamefont {Berezinskii}(1971)}]{Berezinskii:1971}%
  \BibitemOpen
  \bibfield  {author} {\bibinfo {author} {\bibfnamefont {V.}~\bibnamefont
  {Berezinskii}},\ }\href@noop {} {\bibfield  {journal} {\bibinfo  {journal}
  {Soviet Physics {JETP}}\ }\textbf {\bibinfo {volume} {32}},\ \bibinfo {pages}
  {493} (\bibinfo {year} {1971})}\BibitemShut {NoStop}%
\bibitem [{\citenamefont {Kosterlitz}\ and\ \citenamefont
  {Thouless}(1973)}]{Kosterlitz:1973xp}%
  \BibitemOpen
  \bibfield  {author} {\bibinfo {author} {\bibfnamefont {J.~M.}\ \bibnamefont
  {Kosterlitz}}\ and\ \bibinfo {author} {\bibfnamefont {D.~J.}\ \bibnamefont
  {Thouless}},\ }\href@noop {} {\bibfield  {journal} {\bibinfo  {journal} {J.
  Phys.}\ }\textbf {\bibinfo {volume} {C6}},\ \bibinfo {pages} {1181} (\bibinfo
  {year} {1973})}\BibitemShut {NoStop}%
\bibitem [{\citenamefont {Benfatto}\ \emph {et~al.}(2013)\citenamefont
  {Benfatto}, \citenamefont {Castellani},\ and\ \citenamefont
  {Giamarchi}}]{sgrvw1}%
  \BibitemOpen
  \bibfield  {author} {\bibinfo {author} {\bibfnamefont {L.}~\bibnamefont
  {Benfatto}}, \bibinfo {author} {\bibfnamefont {C.}~\bibnamefont
  {Castellani}}, \ and\ \bibinfo {author} {\bibfnamefont {T.}~\bibnamefont
  {Giamarchi}},\ }in\ \href@noop {} {\emph {\bibinfo {booktitle} {40 years of
  {B}eresinskii-{K}osterlitz-{T}houless theory}}},\ \bibinfo {editor} {edited
  by\ \bibinfo {editor} {\bibfnamefont {J.~V.}\ \bibnamefont {Jos{\'e}}}}\
  (\bibinfo  {publisher} {{W}orld {S}cientific},\ \bibinfo {address}
  {Singapore},\ \bibinfo {year} {2013})\ pp.\ \bibinfo {pages} {161--199},\
  \Eprint {http://arxiv.org/abs/1201.2307} {1201.2307} \BibitemShut {NoStop}%
\bibitem [{\citenamefont {Giedt}\ and\ \citenamefont
  {Flamino}(2018)}]{Giedt:2017teu}%
  \BibitemOpen
  \bibfield  {author} {\bibinfo {author} {\bibfnamefont {J.}~\bibnamefont
  {Giedt}}\ and\ \bibinfo {author} {\bibfnamefont {J.}~\bibnamefont
  {Flamino}},\ }\bibfield  {booktitle} {\emph {\bibinfo {booktitle}
  {{Proceedings, 35th International Symposium on Lattice Field Theory (Lattice
  2017): Granada, Spain, June 18-24, 2017}}},\ }\href {\doibase
  10.1051/epjconf/201817514003} {\bibfield  {journal} {\bibinfo  {journal} {EPJ
  Web Conf.}\ }\textbf {\bibinfo {volume} {175}},\ \bibinfo {pages} {14003}
  (\bibinfo {year} {2018})},\ \Eprint {http://arxiv.org/abs/1710.03188}
  {arXiv:1710.03188 [hep-lat]} \BibitemShut {NoStop}%
\bibitem [{\citenamefont {Chui}\ and\ \citenamefont
  {Weeks}(1976)}]{ChuiWeeks76}%
  \BibitemOpen
  \bibfield  {author} {\bibinfo {author} {\bibfnamefont {S.}~\bibnamefont
  {Chui}}\ and\ \bibinfo {author} {\bibfnamefont {J.}~\bibnamefont {Weeks}},\
  }\href@noop {} {\bibfield  {journal} {\bibinfo  {journal} {Phys. Rev.}\
  }\textbf {\bibinfo {volume} {B14}},\ \bibinfo {pages} {4978} (\bibinfo {year}
  {1976})}\BibitemShut {NoStop}%
\bibitem [{\citenamefont {Chui}\ and\ \citenamefont
  {Weeks}(1978)}]{ChuiWeeks78}%
  \BibitemOpen
  \bibfield  {author} {\bibinfo {author} {\bibfnamefont {S.}~\bibnamefont
  {Chui}}\ and\ \bibinfo {author} {\bibfnamefont {J.}~\bibnamefont {Weeks}},\
  }\href@noop {} {\bibfield  {journal} {\bibinfo  {journal} {Phys. Rev. Lett.}\
  }\textbf {\bibinfo {volume} {40}},\ \bibinfo {pages} {733} (\bibinfo {year}
  {1978})}\BibitemShut {NoStop}%
\bibitem [{\citenamefont {Coleman}(1975)}]{Coleman:1974bu}%
  \BibitemOpen
  \bibfield  {author} {\bibinfo {author} {\bibfnamefont {S.~R.}\ \bibnamefont
  {Coleman}},\ }\href {\doibase 10.1103/PhysRevD.11.2088} {\bibfield  {journal}
  {\bibinfo  {journal} {Phys. Rev.}\ }\textbf {\bibinfo {volume} {D11}},\
  \bibinfo {pages} {2088} (\bibinfo {year} {1975})},\ \bibinfo {note}
  {[,128(1974)]}\BibitemShut {NoStop}%
\bibitem [{\citenamefont {Hasenbusch}\ \emph {et~al.}(1994)\citenamefont
  {Hasenbusch}, \citenamefont {Marcu},\ and\ \citenamefont
  {Pinn}}]{Hasenbusch:1994ef}%
  \BibitemOpen
  \bibfield  {author} {\bibinfo {author} {\bibfnamefont {M.}~\bibnamefont
  {Hasenbusch}}, \bibinfo {author} {\bibfnamefont {M.}~\bibnamefont {Marcu}}, \
  and\ \bibinfo {author} {\bibfnamefont {K.}~\bibnamefont {Pinn}},\ }\href
  {\doibase 10.1016/0378-4371(94)00196-0} {\bibfield  {journal} {\bibinfo
  {journal} {Physica}\ }\textbf {\bibinfo {volume} {A211}},\ \bibinfo {pages}
  {255} (\bibinfo {year} {1994})},\ \Eprint
  {http://arxiv.org/abs/hep-lat/9408005} {arXiv:hep-lat/9408005 [hep-lat]}
  \BibitemShut {NoStop}%
\bibitem [{\citenamefont {Duane}\ \emph {et~al.}(1985)\citenamefont {Duane},
  \citenamefont {Kennedy}, \citenamefont {Pendelton},\ and\ \citenamefont
  {Roweth}}]{Duane85}%
  \BibitemOpen
  \bibfield  {author} {\bibinfo {author} {\bibfnamefont {S.}~\bibnamefont
  {Duane}}, \bibinfo {author} {\bibfnamefont {A.}~\bibnamefont {Kennedy}},
  \bibinfo {author} {\bibfnamefont {B.}~\bibnamefont {Pendelton}}, \ and\
  \bibinfo {author} {\bibfnamefont {D.}~\bibnamefont {Roweth}},\ }\href@noop {}
  {\bibfield  {journal} {\bibinfo  {journal} {Phys. Lett.}\ }\textbf {\bibinfo
  {volume} {195}},\ \bibinfo {pages} {216} (\bibinfo {year}
  {1985})}\BibitemShut {NoStop}%
\bibitem [{\citenamefont {Ferreira}\ and\ \citenamefont
  {Toral}(1993)}]{Ferreira93}%
  \BibitemOpen
  \bibfield  {author} {\bibinfo {author} {\bibfnamefont {A.}~\bibnamefont
  {Ferreira}}\ and\ \bibinfo {author} {\bibfnamefont {R.}~\bibnamefont
  {Toral}},\ }\href@noop {} {\bibfield  {journal} {\bibinfo  {journal} {Phys.
  Rev.}\ }\textbf {\bibinfo {volume} {E47}},\ \bibinfo {pages} {3848} (\bibinfo
  {year} {1993})}\BibitemShut {NoStop}%
\bibitem [{\citenamefont {Espriu}\ \emph {et~al.}(1997)\citenamefont {Espriu},
  \citenamefont {Koulovassilopoulos},\ and\ \citenamefont
  {Travesset}}]{Espriu:1997jh}%
  \BibitemOpen
  \bibfield  {author} {\bibinfo {author} {\bibfnamefont {D.}~\bibnamefont
  {Espriu}}, \bibinfo {author} {\bibfnamefont {V.}~\bibnamefont
  {Koulovassilopoulos}}, \ and\ \bibinfo {author} {\bibfnamefont
  {A.}~\bibnamefont {Travesset}},\ }\href {\doibase 10.1103/PhysRevD.56.6885}
  {\bibfield  {journal} {\bibinfo  {journal} {Phys. Rev.}\ }\textbf {\bibinfo
  {volume} {D56}},\ \bibinfo {pages} {6885} (\bibinfo {year} {1997})},\ \Eprint
  {http://arxiv.org/abs/hep-lat/9705027} {arXiv:hep-lat/9705027 [hep-lat]}
  \BibitemShut {NoStop}%
\bibitem [{\citenamefont {Catterall}\ and\ \citenamefont
  {Karamov}(2002)}]{Catterall:2001jg}%
  \BibitemOpen
  \bibfield  {author} {\bibinfo {author} {\bibfnamefont {S.}~\bibnamefont
  {Catterall}}\ and\ \bibinfo {author} {\bibfnamefont {S.}~\bibnamefont
  {Karamov}},\ }\href {\doibase 10.1016/S0370-2693(02)01217-0} {\bibfield
  {journal} {\bibinfo  {journal} {Phys. Lett.}\ }\textbf {\bibinfo {volume}
  {B528}},\ \bibinfo {pages} {301} (\bibinfo {year} {2002})},\ \Eprint
  {http://arxiv.org/abs/hep-lat/0112025} {arXiv:hep-lat/0112025 [hep-lat]}
  \BibitemShut {NoStop}%
\bibitem [{\citenamefont {Kosterlitz}(1974)}]{KosterlitzRG74}%
  \BibitemOpen
  \bibfield  {author} {\bibinfo {author} {\bibfnamefont {J.~M.}\ \bibnamefont
  {Kosterlitz}},\ }\href@noop {} {\bibfield  {journal} {\bibinfo  {journal}
  {{J. Phys.}}\ }\textbf {\bibinfo {volume} {{C7}}},\ \bibinfo {pages} {1046}
  (\bibinfo {year} {1974})}\BibitemShut {NoStop}%
\bibitem [{\citenamefont {Amit}\ \emph {et~al.}(1980)\citenamefont {Amit},
  \citenamefont {Goldschmidt},\ and\ \citenamefont {Grinstein}}]{Amit80}%
  \BibitemOpen
  \bibfield  {author} {\bibinfo {author} {\bibfnamefont {D.~J.}\ \bibnamefont
  {Amit}}, \bibinfo {author} {\bibfnamefont {Y.~Y.}\ \bibnamefont
  {Goldschmidt}}, \ and\ \bibinfo {author} {\bibfnamefont {G.}~\bibnamefont
  {Grinstein}},\ }\href@noop {} {\bibfield  {journal} {\bibinfo  {journal} {{J.
  Phys.}}\ }\textbf {\bibinfo {volume} {{A13}}},\ \bibinfo {pages} {585}
  (\bibinfo {year} {1980})}\BibitemShut {NoStop}%
\bibitem [{\citenamefont {Zinn-Justin}(2002)}]{ZinnJustin}%
  \BibitemOpen
  \bibfield  {author} {\bibinfo {author} {\bibfnamefont {J.}~\bibnamefont
  {Zinn-Justin}},\ }\href@noop {} {\emph {\bibinfo {title} {{Quantum Field
  Theory and Critical Phenomena (4th Edition)}}}}\ (\bibinfo  {publisher}
  {Oxford University Press},\ \bibinfo {address} {New York},\ \bibinfo {year}
  {2002})\BibitemShut {NoStop}%
\bibitem [{\citenamefont {Samuel}(1978)}]{samuel}%
  \BibitemOpen
  \bibfield  {author} {\bibinfo {author} {\bibfnamefont {S.}~\bibnamefont
  {Samuel}},\ }\href@noop {} {\bibfield  {journal} {\bibinfo  {journal}
  {Physical Review}\ }\textbf {\bibinfo {volume} {D18}},\ \bibinfo {pages}
  {1916} (\bibinfo {year} {1978})}\BibitemShut {NoStop}%
\bibitem [{\citenamefont {Mudry}(2014)}]{Mudry}%
  \BibitemOpen
  \bibfield  {author} {\bibinfo {author} {\bibfnamefont {C.}~\bibnamefont
  {Mudry}},\ }\href@noop {} {\emph {\bibinfo {title} {Lecture notes on field
  theory in condensed matter physics}}}\ (\bibinfo  {publisher} {World
  Scientific},\ \bibinfo {address} {New Jersey},\ \bibinfo {year}
  {2014})\BibitemShut {NoStop}%
\bibitem [{\citenamefont {Halperin}\ and\ \citenamefont
  {Nelson}(1979{\natexlab{a}})}]{Hal79}%
  \BibitemOpen
  \bibfield  {author} {\bibinfo {author} {\bibfnamefont {B.~I.}\ \bibnamefont
  {Halperin}}\ and\ \bibinfo {author} {\bibfnamefont {D.~R.}\ \bibnamefont
  {Nelson}},\ }\href@noop {} {\bibfield  {journal} {\bibinfo  {journal}
  {{Journal of Low Temperature Physics}}\ }\textbf {\bibinfo {volume} {36}},\
  \bibinfo {pages} {599} (\bibinfo {year} {1979}{\natexlab{a}})}\BibitemShut
  {NoStop}%
\bibitem [{\citenamefont {{'t Hooft}}(1994)}]{tHooftspell}%
  \BibitemOpen
  \bibfield  {author} {\bibinfo {author} {\bibfnamefont {G.}~\bibnamefont {{'t
  Hooft}}},\ }\href@noop {} {\emph {\bibinfo {title} {Under the spell of the
  gauge principle}}}\ (\bibinfo  {publisher} {World Scientific},\ \bibinfo
  {address} {Singapore},\ \bibinfo {year} {1994})\BibitemShut {NoStop}%
\bibitem [{\citenamefont {Fradkin}(2013)}]{Fradkin}%
  \BibitemOpen
  \bibfield  {author} {\bibinfo {author} {\bibfnamefont {E.}~\bibnamefont
  {Fradkin}},\ }\href@noop {} {\emph {\bibinfo {title} {Field theory of
  condensed matter physics, 2nd edition}}}\ (\bibinfo  {publisher} {Cambridge
  University Press},\ \bibinfo {address} {Cambridge, UK},\ \bibinfo {year}
  {2013})\BibitemShut {NoStop}%
\bibitem [{\citenamefont {Halperin}\ and\ \citenamefont
  {Nelson}(1978)}]{Halp78a}%
  \BibitemOpen
  \bibfield  {author} {\bibinfo {author} {\bibfnamefont {B.}~\bibnamefont
  {Halperin}}\ and\ \bibinfo {author} {\bibfnamefont {D.}~\bibnamefont
  {Nelson}},\ }\href@noop {} {\bibfield  {journal} {\bibinfo  {journal} {Phys.
  Rev. Lett.}\ }\textbf {\bibinfo {volume} {41}},\ \bibinfo {pages} {121}
  (\bibinfo {year} {1978})}\BibitemShut {NoStop}%
\bibitem [{\citenamefont {Halperin}\ and\ \citenamefont
  {Nelson}(1979{\natexlab{b}})}]{Halp79a}%
  \BibitemOpen
  \bibfield  {author} {\bibinfo {author} {\bibfnamefont {B.}~\bibnamefont
  {Halperin}}\ and\ \bibinfo {author} {\bibfnamefont {D.}~\bibnamefont
  {Nelson}},\ }\href@noop {} {\bibfield  {journal} {\bibinfo  {journal} {Phys.
  Rev.}\ }\textbf {\bibinfo {volume} {B19}},\ \bibinfo {pages} {2457} (\bibinfo
  {year} {1979}{\natexlab{b}})}\BibitemShut {NoStop}%
\bibitem [{\citenamefont {Young}(1979)}]{Young79}%
  \BibitemOpen
  \bibfield  {author} {\bibinfo {author} {\bibfnamefont {A.}~\bibnamefont
  {Young}},\ }\href@noop {} {\bibfield  {journal} {\bibinfo  {journal} {Phys.
  Rev.}\ }\textbf {\bibinfo {volume} {B19}},\ \bibinfo {pages} {1855} (\bibinfo
  {year} {1979})}\BibitemShut {NoStop}%
\bibitem [{\citenamefont {Nelson}(1978)}]{Nelson78}%
  \BibitemOpen
  \bibfield  {author} {\bibinfo {author} {\bibfnamefont {D.}~\bibnamefont
  {Nelson}},\ }\href@noop {} {\bibfield  {journal} {\bibinfo  {journal} {Phys.
  Rev.}\ }\textbf {\bibinfo {volume} {B18}},\ \bibinfo {pages} {2318} (\bibinfo
  {year} {1978})}\BibitemShut {NoStop}%
\bibitem [{\citenamefont {Pordes}\ and\ \citenamefont {{et al.}}(2007)}]{osg1}%
  \BibitemOpen
  \bibfield  {author} {\bibinfo {author} {\bibfnamefont {R.}~\bibnamefont
  {Pordes}}\ and\ \bibinfo {author} {\bibnamefont {{et al.}}},\ }\href
  {\doibase 10.1088/1742-6596/78/1/012057} {\bibfield  {journal} {\bibinfo
  {journal} {J. Phys. Conf. Ser.}\ }\textbf {\bibinfo {volume} {78}},\ \bibinfo
  {pages} {012057} (\bibinfo {year} {2007})}\BibitemShut {NoStop}%
\bibitem [{\citenamefont {Sfiligoi}\ \emph {et~al.}(2009)\citenamefont
  {Sfiligoi}, \citenamefont {Bradley}, \citenamefont {Holzman}, \citenamefont
  {Mhashilkar}, \citenamefont {Padhi},\ and\ \citenamefont {Wurthwein}}]{osg2}%
  \BibitemOpen
  \bibfield  {author} {\bibinfo {author} {\bibfnamefont {I.}~\bibnamefont
  {Sfiligoi}}, \bibinfo {author} {\bibfnamefont {D.~C.}\ \bibnamefont
  {Bradley}}, \bibinfo {author} {\bibfnamefont {B.}~\bibnamefont {Holzman}},
  \bibinfo {author} {\bibfnamefont {P.}~\bibnamefont {Mhashilkar}}, \bibinfo
  {author} {\bibfnamefont {S.}~\bibnamefont {Padhi}}, \ and\ \bibinfo {author}
  {\bibfnamefont {F.}~\bibnamefont {Wurthwein}},\ }\href {\doibase
  10.1109/CSIE.2009.950} {\bibfield  {journal} {\bibinfo  {journal} {2009 WRI
  World Congress on Computer Science and Information Engineering}\ }\textbf
  {\bibinfo {volume} {2}},\ \bibinfo {pages} {428} (\bibinfo {year}
  {2009})}\BibitemShut {NoStop}%
\bibitem [{\citenamefont {Towns}\ \emph {et~al.}(2014)\citenamefont {Towns},
  \citenamefont {Cockerill}, \citenamefont {Dahan}, \citenamefont {Foster},
  \citenamefont {Gaither}, \citenamefont {Grimshaw}, \citenamefont {Hazlewood},
  \citenamefont {Lathrop}, \citenamefont {Lifka}, \citenamefont {Peterson},
  \citenamefont {Roskies}, \citenamefont {Scott},\ and\ \citenamefont
  {Wilkins-Diehr}}]{xsede}%
  \BibitemOpen
  \bibfield  {author} {\bibinfo {author} {\bibfnamefont {J.}~\bibnamefont
  {Towns}}, \bibinfo {author} {\bibfnamefont {T.}~\bibnamefont {Cockerill}},
  \bibinfo {author} {\bibfnamefont {M.}~\bibnamefont {Dahan}}, \bibinfo
  {author} {\bibfnamefont {I.}~\bibnamefont {Foster}}, \bibinfo {author}
  {\bibfnamefont {K.}~\bibnamefont {Gaither}}, \bibinfo {author} {\bibfnamefont
  {A.}~\bibnamefont {Grimshaw}}, \bibinfo {author} {\bibfnamefont
  {V.}~\bibnamefont {Hazlewood}}, \bibinfo {author} {\bibfnamefont
  {S.}~\bibnamefont {Lathrop}}, \bibinfo {author} {\bibfnamefont
  {D.}~\bibnamefont {Lifka}}, \bibinfo {author} {\bibfnamefont {G.~D.}\
  \bibnamefont {Peterson}}, \bibinfo {author} {\bibfnamefont {R.}~\bibnamefont
  {Roskies}}, \bibinfo {author} {\bibfnamefont {J.~R.}\ \bibnamefont {Scott}},
  \ and\ \bibinfo {author} {\bibfnamefont {N.}~\bibnamefont {Wilkins-Diehr}},\
  }\href {\doibase doi.ieeecomputersociety.org/10.1109/MCSE.2014.80} {\bibfield
   {journal} {\bibinfo  {journal} {Computing in Science \& Engineering}\
  }\textbf {\bibinfo {volume} {16}},\ \bibinfo {pages} {62} (\bibinfo {year}
  {2014})}\BibitemShut {NoStop}%
\bibitem [{\citenamefont {Gradshteyn}\ and\ \citenamefont
  {Ryzhik}(1994)}]{GR94}%
  \BibitemOpen
  \bibfield  {author} {\bibinfo {author} {\bibfnamefont {I.~S.}\ \bibnamefont
  {Gradshteyn}}\ and\ \bibinfo {author} {\bibfnamefont {I.~M.}\ \bibnamefont
  {Ryzhik}},\ }\href@noop {} {\emph {\bibinfo {title} {{Table of Integrals,
  Series \& Products, Fifth Edition}}}},\ edited by\ \bibinfo {editor}
  {\bibfnamefont {A.}~\bibnamefont {Jeffrey}}\ (\bibinfo  {publisher} {Academic
  Press},\ \bibinfo {address} {San Diego, California},\ \bibinfo {year}
  {1994})\BibitemShut {NoStop}%
\end{thebibliography}%
\bibliographystyle{apsrev4-1}

\end{document}